\documentclass[a4paper,fleqn,usenatbib,useAMS]{mnras}

\usepackage{graphicx}	
\usepackage{multicol}        
\usepackage{pdflscape}	
\RequirePackage{bm,amsfonts,hhline,amsmath,amssymb,microtype,float,eurosym,latexsym,epsf,mathtools,
cuted,times,makecell,array,natbib}
\usepackage[font={footnotesize,it}]{caption}
\usepackage[caption = false]{subfig}
\usepackage{epstopdf}

\usepackage{hyperref}

\newcommand*\DAlambert{\mathop{}\!\mathbin\Box}
\usepackage[switch]{lineno}

\begin{document}

\title[Anisotropic strange stars]{Exploring physical features of anisotropic strange stars\\ beyond standard maximum mass limit in $f\left(R,\mathcal{T}\right)$ gravity}

\author[Deb, Ketov, Maurya, Khlopov, Moraes \& Ray]{Debabrata Deb$^{1}$\thanks{E-mail: ddeb.rs2016@physics.iiests.ac.in}, Sergei V. Ketov$^{2}$\thanks{E-mail: ketov@tmu.ac.jp}, S.K. Maurya$^{3}$\thanks{E-mail: sunil@unizwa.edu.om}, Maxim Khlopov$^{4}$\thanks{E-mail: khlopov@apc.in2p3.fr},\newauthor P.H.R.S. Moraes$^{5}$\thanks{E-mail: moraes.phrs@gmail.com} and Saibal Ray$^{6}$\thanks{E-mail: saibal@associates.iucaa.in}\\ $^{1}$Department of Physics, Indian Institute of Engineering Science and Technology, Shibpur, Howrah, West Bengal, 711103, India\\ $^{2}$ Research School of High-Energy Physics, Tomsk Polytechnic University, 2a Lenin Avenue, Tomsk 634050, Russian Federation \\ \& Department of Physics, Tokyo Metropolitan University, Minami-ohsawa 1-1, Hachioji-shi, Tokyo 192-0397, Japan \\ \& Institute for Theoretical Physics, Vienna University of Technology, Wiedner Hauptstrasse 8-10/136, A-1040 Vienna, Austria \\ \& Kavli Institute for the Physics and Mathematics of the Universe (IPMU), The University of Tokyo, Chiba 277-8568, Japan\\ $^{3}$ Department of Mathematical and Physical Sciences, College of Arts and Science, University of Nizwa, Nizwa, Sultanate of Oman\\ $^{4}$ Institute of Physics, Southern Federal University, 194 Stachki, Rostov-on-Don 344090, Russian Federation \\ \& APC Laboratory 10, rue Alice Domon et L{\'e}onie Duquet, 75205 Paris Cedex 13, France\\ \& Center for Cosmopartcile Physics Cosmion and National Research Nuclear University MEPHI (Moscow State Engineering Physics Institute),\\
~~~ Kashirskoe Shosse 31, Moscow 115409, Russia \\ $^{5}$ ITA - Instituto Tecnol\'ogico de Aeron\'autica - Departamento de F\'isica, 12228-900, S\~ao Jos\'e dos Campos, S\~ao Paulo, Brazil \\ $^{6}$ Department of Physics, Government College of Engineering and Ceramic Technology, Kolkata 700010, West Bengal, India\\ \& Department of Natural Sciences, Maulana Abul Kalam Azad University of Technology, Haringhata 741249, West Bengal, India}

\date{Accepted . Received ; in original form }

\pagerange{\pageref{firstpage}--\pageref{lastpage}} \pubyear{2018}

\maketitle

\begin{abstract}
We study a specific model of anisotropic strange stars in the modified $f\left(R,\mathcal{T}\right)$-type gravity by deriving solutions to the modified Einstein field equations representing a spherically symmetric anisotropic stellar object. We take a standard assumption that $f(R,\mathcal{T})=R+2\chi\mathcal{T}$, where $R$ is Ricci scalar, $\mathcal{T}$ is the trace of the energy-momentum tensor of matter, and $\chi$ is a coupling constant. To obtain our solution to the modified Einstein equations, we successfully apply the `embedding class 1' techniques. We also consider the case when the strange quark matter (SQM) distribution is governed by the simplified MIT bag model equation of state given by $p_r=\frac{1}{3}\left(\rho-4B\right)$, where $B$ is bag constant. We calculate the radius of the strange star candidates by directly solving the modified TOV equation with the observed values of the mass and some parametric values of $B$ and $\chi$. The physical acceptability of our solutions is verified by performing several physical tests. Interestingly, besides the SQM, another type of matter distribution originates due to the effect of coupling between the matter and curvature terms in the $f\left(R,\mathcal{T}\right)$ gravity theory. Our study shows that with decreasing the value of $\chi$, the stellar systems under investigations become gradually massive and larger in size, turning them into less dense compact objects. It also reveals that for $\chi<0$ the $f\left(R,\mathcal{T}\right)$ gravity emerges as a suitable theory for explaining the observed massive stellar objects like massive pulsars, super-Chandrasekhar stars and magnetars, etc., which remain obscure in the standard framework of General Relativity (GR).
\end{abstract}


\section{Introduction}\label{sec:intro} 
It is well established by several independent observations of the distant indicators and standard candles that we are living in the age of the accelerated expansion of the Universe. This is certainly not the outcome of the ordinary matter considered as a perfect fluid, which was supposed to be the source in the cosmological Friedmann equations~\citep{Riess1998,Perlmutter1999,Riess2004}. As a result of this evidence, an explanation of the evolution of the large scale structures of the galaxies as well as Universe is facing a serious challenge. In addition, the observations of the cosmic microwave background radiation (CMBR) anisotropies~\citep{Spergel2003}, the Lyman-$\alpha$ forest power spectrum from the Sloan Digital Sky Survey~\citep{McDonald2006} and investigation of the high-energy-physics-motivated models of dark energy with the weak lensing data~\citep{Schimd2007} clearly demonstrate an acceleration in the Hubble fluid.
  
To overcome the apparent discrepancy between the critical density of the spatially flat Universe and the amount of the observed luminous matter, an assumption about the presence of a non-standard cosmic fluid having negative pressure is essential. In the large scale structure, this cosmic fluid is not clustered. One can explain the observed accelerating Universe due to a positive cosmological constant (contributing about $70\%$ to the total energy of the Universe), which is consistent with the observed acceleration in the Hubble fluid. The main part of the rest of the energy budget is supposed to be due the presence of dark matter (contributing about $25\%$) that is clustered in the large scale structure. This is known as the ${\Lambda}$CDM model~\citep{Bahcall1999}. However, when the dark energy (and hence the cosmological constant) is identified with the vacuum (ground state) energy as proposed by~\citet{Zel'dovich1967a,Zel'dovich1967b}, its observed value in the ${\Lambda}$CDM model is in huge discrepancy with its expected theoretical value in quantum gravity~\citep{Weinberg1989} by the $120$ orders of magnitude. This `cosmological constant problem' is the most fundamental problem in the modern age of cosmology. Hence, it is essential to find a better explanation for the dark matter and dark energy. As, the standard General Relativity (GR) faces difficulties in explaining the universe at large scales beyond the Solar system, as well as at large energies, it seems natural to modify the standard Einstein gravity framework towards the better understanding of gravity at a very large and very small scales. It is worth mentioning that GR hardly can explain the observations of massive pulsars~\citep{Demorest2010,Antoniadis2013} and white dwarfs~\citep{Howell2006,Kepler2007,Scalzo2010} which have masses beyond the standard maximum mass limit, i.e., $1.44~M_{\odot}$ (the Chandrasekhar mass Limit), satisfactorily.
 
In the so-called alternative gravity theories, one starts with extended or modified Einstein equations 
that may explain the known shortcomings of GR without including `dark' components~\citep{Capozziello2002,Capozziello2003a,Capozziello2003b,Carroll2004,Nojiri2007,Capozziello2010,Capozziello2011,Capozziello2012a,CruzDombriz2012}. For instance, the Hubble diagram can be successfully reproduced from the SNela surveys~\citep{Capozziello2003b,Demianski2006}, as well as the observed CMBR anisotropies can be justified~\citep{Perrotta1999,Hwang2001}. Still, one should not ignore the fact that Einstein GR is the extremely successful theory, having a lot of experimental supports including many precision measurements. The conservative (Einstein GR based) ${\Lambda}$CDM model is still phenomenologically acceptable and has the strong experimental evidence due to the measured ratio between the pressure and energy density in the effective `ideal fluid' equation of state, which is very close to $(-1)$, which corresponds to the cosmological constant. Giving up some basic and established principles for `explaining' a particular unknown phenomenon should be exercised with great care, because it may easily result into a failure due to the `re-explanation' by a new theory to the already known and explained phenomena by the standard theory, not to mention the famous ``Occam razor" conceptual principle of thinking. Hence, the more conservative (or small) a  GR modification, is the better in general.

The standard Einstein gravity theory currently fails to explain the exotic astrophysical stellar structures. This may be resolved by studying relativistic stellar objects in the extended theory of gravity proposed in the refs.~\citep{Capozziello2011a,Capozziello2012b}. The observational data from local tests of gravity may also be consistent with some alternative theories of gravity \citep{Nojiri2011}. There exist many other alternative gravity theories, such as the $f\left(R\right)$ gravity~\citep{Carroll2004,Allemandi2005,Nojiri2007,Bertolami2007,DeFelice2012}, the $f\left(\mathbb{T}\right)$ gravity~\citep{Bengochea2009,Linder2010,Bohmer2011,Cai2016}, the $f\left(G\right)$ gravity~\citep{Bamba2010a,Bamba2010b,Rodrigues2014}, the $f\left(R,G\right)$ gravity~\citep{Nojiri2005,Laurentis2015}, Brans-Dicke (BD) gravity~\citep{bhattacharya/2015,avilez/2014}, etc., where $R$, $G$ and $\mathbb{T}$ are the Ricci scalar, Gauss-Bonnet scalar and torsion scalar, respectively. 

It has been investigated that $f\left(R\right)$ gravity theories largely failed to be consistent with the Solar system tests~\citep{Erickcek2006,Chiba2007,Capozziello2007}. Again, in recent times $f\left(R\right)$ gravity theory also faced stringent constraint to explain the galactic scale, CMBR tests and the strong lensing regime~\citep{Dolgov2003,Chiba2003,Olmo2005,Yang2007,Dossett2014,Campigotto2017,Xua2018}. In this connection, it is to note that the $f(R)$ gravity theory actually should not be in the list of the alternative gravity theories as mentioned above, because they are known to be (classically) equivalent to the scalar-tensor gravity theories, i.e. the Einstein GR with a real scalar field whose scalar potential is in dual correspondence to $f$-function~\citep{Fujii2003,Ketov2013}. This also applies to the extensions of $f(R)$ gravity theories in supergravity~\citep{Ketov2010,Ketov2011,Ketov2012,Ketov2013,Addazi2017} and braneworld~\citep{Nakada2016}.

Yet another class of the alternative gravity theories, extending the generalization of the $f\left(R\right)$ gravity theories, was introduced in~\citep{harko2011} by employing the arbitrary function $f\left(R,\mathcal{T}\right)$ in the gravitational Lagrangian, where $\mathcal{T}$ is the trace of the energy-momentum tensor of matter. The idea of Harko and collaborators~\citep{harko2011} when they introduced the $f(R,\mathcal{T})$ gravity was to check if new material, rather than particularly geometrical terms, were able to evade the troubles $f(R)$ gravity faces. This $f\left(R,\mathcal{T}\right)$ gravity theory was applied to cosmology~\citep{Jamil2012,Shabani2013,Shabani2014,Momeni2015,Zaregonbadi2016,Shabani2017a,Shabani2017b} as well as to stellar astrophysics~\citep{sharif2014,noureen2015,noureen2015b,noureen2015c,zubair2015a,zubair2015b,Ahmed2015,Das2016,Deb2018a,Deb2018b}. 
A solution to Tolman-Oppenheimer-Volkoff (TOV) equation for an isotropic spherically symmetric compact stellar system was found by~\citet{Moraes2016} whereas a gravastar model in $f\left(R,\mathcal{T}\right)$ gravity can be found in the work by~\citet{Das2017}.

It is important to realize that any alternative theory of gravity gives up some of the minimal assumptions of GR. To consider quantum effects, such as particle production~\citep{Harko2014}, or exotic imperfect fluids,~\citet{harko2011} introduce the dependence on $T$ in the gravitational Lagrangian. Their study~\citep{harko2011} reveals an extra acceleration due to the effect of the coupling between the geometrical terms and matter, which forces particles to follow a non-geodesic path leaving the covariant derivative of the matter energy-momentum tensor to be non-vanishing, i.e., ${\nabla}_{\mu} {T}^{\mu\nu} \neq 0$~\citep{harko2011,barrientos2014}. In fact, in the point of view of cosmology, the creation (or destruction) of matter during the evolution of the universe recommends the possibility of non-conservation of the energy-momentum tensor~\citep{kumar2015,Singh2016}. The investigation by~\citet{Harko2014} presented a detailed discussion on this topic from the perspective of thermodynamics. One may consult the literature~\citep{Harko2010,Harko2013} for the better understanding of the non-conservative energy-momentum theories. Note that recent studies have revealed the dependency of the cosmic acceleration with the energy-momentum tensor~\citep{Shabani2017c,Shabani2017d,Josset2017}. 

Now, the creation (or destruction) of the matter is happening due to the process occurring in the quantum level which is suitable in the universal scale, whereas in the astrophysical scenario this should not be the case for the static analysis.~\citet{SC2013} in his study resolved this issue with a specific choice of the $f\left(R,\mathcal{T}\right)$ function given by $f\left(R,\mathcal{T}\right)=R+h\left(T\right)$, and showed that another type of fluid originates due to the coupling between the matter and geometry, which turns the stellar system into a non-interacting two-fluid system. The study~\citep{SC2013} importantly features the conservation of the effective energy-momentum tensor with the geodesic motion of particles, in the framework of $\left(R,\mathcal{T}\right)$ gravity theory. Further investigations by~\citet{Shabani2014} also revealed that the $f\left(R,\mathcal{T}\right)$ gravity theory has passed the Solar System test. Also, it has successfully explained the deviation to the usual geodesic equation~\citep{Baffou2017a}, and the gravitational lensing test~\citep{Ahmed2015}. Recent study by~\citet{Zaregonbadi2016a} reveals that $f\left(R,\mathcal{T}\right)$ gravity theory can explain the dark matter galactic effects suitably. More recently, it was shown~\citep{Baffou2017b,Nagpal2018} that the models inspired by the modified $\left(R,\mathcal{T}\right)$ gravity are in good agreement with observations, namely, the Union 2.1 compilation data set SNeIa, the observational Hubble data set $H(z)$, the joint data set $H(z)$+ SNeIa and $H(z)$+ SNeIa + BAO and the Baryon Acoustic Oscillation data BAO. In addition, cosmological dynamics in the unimodular $\left(R,\mathcal{T}\right)$ gravity is consistent with the Planck 2015 observational data \citep{Rajabi2017}.

The effects of anisotropy on the spherically symmetric compact stellar systems in GR were studied in the literature~\citep{Ivanov2002,SM2003,MH2003,Usov2004,Varela2010,Rahaman2010,Rahaman2011,Rahaman2012,Kalam2012,Deb2017,Shee2016,Maurya2016,Maurya2017}. The anisotropy of the system implies that the radial component of the pressure, $p_r(r)$ differs from the angular component, ${p_{\theta}}(r) = {p_{\phi}}(r) \equiv {p_{t}}(r)$. The condition ${p_{\theta}}(r) = {p_{\phi}}(r)$ is the direct result of the spherical symmetry. It is also important to note that the scalar field with a non-zero spatial gradient creates anisotropy in the pressure of a physical system. The main reason for arising anisotropy in a $f\left(R,\mathcal{T}\right)$ gravity theory model could be the anisotropic nature of the non-interacting two fluid system. Also, it should be stressed that from a quantum perspective the $\mathcal{T}$-dependent Lagrangian may be related to the creation of particles which may naturally describe the existence of bulk viscosity and other ``imperfections'' in the referred fluid.

Curved spacetimes, such as those in the vicinity of compact astrophysical objects, can also be considered from the viewpoint of their embedding into higher dimensions of the flat geometry. In 1850, Riemann in his thesis introduced the concept of Riemannian geometry which studies the geometric objects intrinsically. Immediately, the thesis arose a question whether a Riemannian manifold can be represented as a sub-manifold into a higher dimensional Euclidean space which is later widely known as the isometric embedding problem.~\citet{Schlaefli1871} conjectured that a Riemannian manifold, having an analytic metric and a positively defined signature can be embedded locally and isometrically into the higher dimensional Euclidean space. Later,~\citet{Janet1926},~\citet{Cartan1927}, and~\citet{Burstin1931} proved the conjecture and shown that an $n$-dimensional Riemannian space~$V^n$ can be locally and isometrically embedded into an $N=n\left(n+1\right)/2$~dimensional pseudo-Euclidean space. In the symmetric cases, an $n$-dimensional geometry requires at least the $(n + 1)$ dimensional pseudo-Euclidean space for its embedding. The embedding geometry (the pseudo-Euclidean space) must necessarily have higher numbers of positive and negative eigenvalues than those of the embedded space. The so-called ``embedding class" is the number of extra dimensions needed for the minimal embedding. The embedding class $\left(p\right)$ of any four-dimensional Lorentzian geometry must lie between $p=1$ to $p=6$. The constant curvature spaces belong to class 1, whereas the Schwarzschild solution is in class 2. Along with the other classification schemes of Petrov type, the invariant classification scheme can be based on the embedding classes of GR solutions~\citep{Stephani2003}. For further study one may consult referred literature~\citep{Barnes1974,Kumar2010,Barnes2011,Leon2015,Akbar2017,Abbas2018,Kuhfittig2018a,Kuhfittig2018b} where authors have explicitly applied and discussed the effects of the technique of embedding of lower dimensional Riemannian space into the higher dimensional pseudo-Euclidean space in the framework of GR and modified gravity.

The ultra-dense compact astrophysical objects made up out of $(u)$, $(d)$ and $(s)$ quark matter are known as strange stars~(SS).~\citet{Bodmer1971} and~\citet{Witten1984} conjectured the possible existence of SS with the strange quark matter (SQM) as the absolute ground state of the strongly interacting matter. This new type of compact stars has drawn the attention of many theorists~\citep{Baym1976,Haensel1986,Alcock1986,Drago1996}. The situation drastically changed after a large amount of observational data was collected by using the new generation $\gamma$ and X-ray satellites. It was shown that $Her~X-1$ and $4U~1820-30$ are the possible candidates for strange stars~\citep{Li1995,Bombaci1997,Dey1998}.~\citet{Demorest2010} determined the mass of the SS candidate $PSR~J1614+2230$ by using Shapiro delay and predicted that such the high mass stellar object can only be justified by using MIT bag equation state (EOS).~\citet{Gangopadhyay2013} studied $12$ possible candidates of strange stars and predicted their radii by using the observed mass of the compact stellar objects.

In this paper, in order to obtain a solution to the modified Einstein field equations in the framework of $f\left(R,\mathcal{T}\right)$ theory of gravity, we use the embedding class one technique, via embedding the interior 
four-dimensional spacetime into a five-dimensional flat Euclidean space. Our paper is organized as follows: 
in Section~\ref{sec1} we provide a basic formulation of our $f\left(R,\mathcal{T}\right)$ gravity theory. The stellar equations of the spherically symmetric anisotropic system inspired by the modified $f\left(R,\mathcal{T}\right)$ theory of gravity, are given in Section~\ref{sec2}, and their general solutions, by employing the embedding class one formalism, are given in Section~\ref{sec3}. By using matching boundary conditions in Section~\ref{sec4}, we explore various features of our stellar model in Section~\ref{sec5} which are physically plausible and interesting. We conclude our study with a brief discussion of the outcome in Section~\ref{sec6}.

\section{Basic mathematical formulation of $f(R,\mathcal{T})$ Theory}\label{sec1} 
The action of the $f(R,\mathcal{T})$ theory~\citep{harko2011} is 
 \begin{equation}\label{1.1}
S=\frac{1}{16\pi}\int d^{4}xf(R,\mathcal{T})\sqrt{-g}+\int
d^{4}x\mathcal{L}_m\sqrt{-g},
\end{equation}
where $f(R,\mathcal{T})$ is a general function of the Ricci scalar $R$
and the trace of the energy-momentum tensor $\mathcal{T}$, $g$ is
the determinant of the metric $g_{\mu\nu}$ and $\mathcal{L}_m$ is 
the matter Lagrangian density. In the further study, we adopt the geometrical units $G=c=1$.

Varying the action (\ref{1.1}) with respect to the metric $g_{\mu\nu}$ yields
the field equation 
\begin{eqnarray}\label{1.2}
&\qquad\hspace{-1cm} f_R (R,\mathcal{T}) R_{\mu\nu} - \frac{1}{2} f(R,\mathcal{T}) g_{\mu\nu} 
+ (g_{\mu\nu}\DAlambert - \nabla_{\mu} \nabla_{\nu}) f_R (R,\mathcal{T})\nonumber \\
&\qquad\hspace{1cm} = 8\pi T_{\mu\nu} - f_\mathcal{T}(R,\mathcal{T}) T_{\mu\nu} -
f_\mathcal{T}(R,\mathcal{T})\Theta_{\mu\nu},\end{eqnarray}
where ${f_R (R,\mathcal{T})= \partial f(R,\mathcal{T})/\partial R}$,
${f_\mathcal{T}(R,\mathcal{T})=\partial f(R,\mathcal{T})/\partial \mathcal{T}}$, 
${\DAlambert \equiv\partial_{\mu}(\sqrt{-g} g^{\mu\nu} \partial_{\nu})/\sqrt{-g}}$ is the D'Alambert operator,
$R_{\mu\nu}$ is the Ricci tensor, $\nabla_\mu$ represents the
covariant derivative associated with the Levi-Civita connection of $g_{\mu\nu}$, $\Theta_{\mu\nu}=
g^{\alpha\beta}\delta T_{\alpha\beta}/\delta g^{\mu\nu}$ and $T_{\mu\nu}$ is the
stress-energy tensor defined as $T_{\mu\nu}=g_{\mu\nu}\mathcal{L}_m-2\partial\mathcal{L}_m/\partial
g^{\mu\nu}$~\citep{Landau2002}.

The covariant divergence of (\ref{1.2}) yields~\citep{barrientos2014}
\begin{eqnarray}\label{1.3}
&\qquad\hspace{-2cm}\nabla^{\mu}T_{\mu\nu}=\frac{f_\mathcal{T}(R,\mathcal{T})}{8\pi -f_\mathcal{T}(R,\mathcal{T})}[(T_{\mu\nu}+\Theta_{\mu\nu})\nabla^{\mu}\ln f_\mathcal{T}(R,\mathcal{T}) \nonumber \\
&\qquad\hspace{2cm} +\nabla^{\mu}\Theta_{\mu\nu}-(1/2)g_{\mu\nu}\nabla^{\mu}\mathcal{T}],
\end{eqnarray}
where equation (\ref{1.3}) features that the energy-momentum tensor in $f(R,T)$ gravity is not conserved, as in other theories of gravity \citep{zhao/2012,yu/2018}.
 
As $\mathcal{T}$-dependence presumably invites quantum effects, such as the production of particles or the consideration of exotic imperfect fluids, in the present study we are considering the energy-momentum tensor for the anisotropic fluid is given as
\begin{equation}\label{1.4}
T_{\mu\nu}=(\rho+{p_t})u_\mu u_\nu-{p_t}g_{\mu\nu}+\left({p_r}-{p_t}\right)v_\mu v_\nu,
\end{equation}
where ${v_{\mu}}$ and ${u_{\nu}}$ are the radial-four vectors and four velocity vectors, respectively. Here, $\rho$ represents matter density, whereas $p_r$ and $p_t$ are the radial and tangential pressures, respectively. In the present study, we choose $\mathcal{L}_m=-\mathcal{P}$, where $\mathcal{P}=\frac{1}{3}\left(p_r+2p_t\right)$ and we have $\Theta_{\mu\nu}=-2T_{\mu\nu}-\mathcal{P} g_{\mu\nu}$.

Following~\citet{harko2011} we assume the simplified and linear functional form of $f(R,\mathcal{T})$ as $f(R,\mathcal{T})=R+2\chi\mathcal{T}$, where $\chi$ is a constant. This specific functional form has been used successfully in many $f(R,\mathcal{T})$ gravity models~\citep{singh2015,moraes2014b,moraes2015a,moraes2016b,moraes2017,reddy2013b,kumar2015,shamir2015,Fayaz2016}.

Now substituting the assumed form of $f(R,\mathcal{T})$ in Eq.~(\ref{1.2}) we have 
\begin{eqnarray}\label{1.5}
G_{\mu\nu}=8\pi T_{\mu\nu}+\chi \mathcal{T}g_{\mu\nu}+2\chi(T_{\mu\nu}+\mathcal{P} g_{\mu\nu})=8\pi T^{eff}_{\mu\nu},
\end{eqnarray}
 where $G_{\mu\nu}$ is the usual Einstein tensor and $T^{eff}_{\mu\nu}$ is the effective energy-momentum tensor given as
   \begin{eqnarray}\label{1.5a}
   T^{eff}_{\mu\nu}=T_{\mu\nu}+\frac{\chi}{8\pi} \mathcal{T}g_{\mu\nu}+\frac{\chi}{4\pi}(T_{\mu\nu}+\mathcal{P} g_{\mu\nu}).
   \end{eqnarray}
 
 Putting $\chi=0$ in Eq.~(\ref{1.5}) reproduces  the standard GR results. For $f(R,\mathcal{T})=R+2\chi\mathcal{T}$ Eq.~(\ref{1.3}) yields 
 \begin{eqnarray}\label{1.6}
 \left(4\pi+\chi\right)\nabla^{\mu}T_{\mu\nu}=-\frac{1}{2}\chi\left[g_{\mu\nu}\nabla^{\mu}\mathcal{T}+2\,\nabla^{\mu}(\mathcal{P} g_{\mu\nu})\right].
 \end{eqnarray}

\section{Einstein's field equations in $f\left(R,\mathcal{T}\right)$ gravity}\label{sec2}
We assume the interior spacetime of the spherically symmetric and static stellar configuration is described by the metric 
 \begin{equation}\label{2.1}
ds^2=e^{\nu(r)}dt^2-e^{\lambda(r)}dr^2-r^2(d\theta^2+\sin^2\theta d\phi^2),
\end{equation}
 where $\nu$ and $\lambda$ are the metric potentials and functions of the radial coordinate, $r$ only.
 
 Now using Eqs.~(\ref{1.4}), (\ref{1.5}) and (\ref{2.1}) we have the Einstein field equations explicitly for $f\left(R,\mathcal{T}\right)$ gravity as
 {\small{\begin{eqnarray}\label{2.2}
 &\qquad\hspace{-1.2cm} {{\rm e}^{-\lambda}} \left( {\frac {\lambda^{{\prime}}}{r}}-\frac{1}{{r}^{2}}
 \right) +\frac{1}{{r}^{2}}= 8\,\pi \,\rho+\frac{\chi}{3}\,\left( 9\,\rho-p_{{r}}-2\,p_{{t}} \right) =8\pi{\rho}^{{\it eff}},\\ \label{2.3}
&\qquad\hspace{-1.2cm} {{\rm e}^{-\lambda}} \left( {\frac {\nu^{{\prime}}}{r}}+\frac{1}{{r}^{2}} \right) -\frac{1}{{r}^{2}}=8\,\pi \,p_{{r}}-\frac{\chi}{3}\, \left( 3\,\rho-7\,p_{{r}}-2\,p_{{t}}
 \right) =8\pi{p^{{\it eff}}_r},\\ \label{2.4}
&\qquad\hspace{-4cm} \frac{{{\rm e}^{-\lambda}}}{2} \left( \nu^{{\prime\prime}}+\frac{1}{2}\,{{\nu}^{{\prime}}}^{2}+{\frac {\nu^{{\prime}}-\lambda^{{\prime}}}{r}}-\frac{1}{2}\,\nu^{{\prime}}\lambda^{{\prime}} \right) \nonumber \\
&\qquad\hspace{1.1cm} =8\,\pi \,p_{{t}}-\frac{\chi}{3}\, \left( 3\,\rho-p_{{r}}-8\,p_{{t}} \right) =8\pi{p^{{\it eff}}_t},
 \end{eqnarray}}}
where the prime $(\prime)$ denotes the differentiation with respect to the radial coordinate. Here, ${\rho}^{{\it eff}}$,~${p^{{\it eff}}_r}$~and~${p^{{\it eff}}_t}$ represent the effective density, effective radial pressure and effective tangential pressure of the compact stellar system as follows:
\begin{eqnarray}\label{effeq1}
{{\rho}^{{\it eff}}}=\rho+{\frac {\chi\,}{24\pi}}\left( 9\,\rho-p_{{r}}-2\,p_{{t}} \right),\\ \label{effeq2}
{{p}^{{\it eff}}_r}=p_{{r}}-{\frac {\chi\,}{24\pi}}\left( 3\,\rho-7\,p_{{r}}-2\,p_{{t}} \right),\\ \label{effeq3}
{{p}^{{\it eff}}_t}=p_{{t}}-{\frac {\chi\,}{24 \pi }}\left( 3\,\rho-p_{{r}}-8\,p_{{t}} \right) .
\end{eqnarray}
  
To solve Eqs.~(\ref{2.2})-(\ref{2.4}), we assume the equation of state (EOS) inside the stellar system to be governed by the well known MIT bag EOS~\citep{Chodos1974}. To include all the corrections of energy and pressure functions of SQM in the MIT bag model, an {\it ad hoc} bag function has been introduced. By considering that the quarks are massless and non-interacting in this simplified bag model the quark pressure, ${p_r}$ can be defined as
 \begin{equation}
{p_r}={\sum_{f=u,d,s}}{p^f}-{B}, \label{eqeos1}
\end{equation}
 where ${p^f}$ is the individual pressure of the each quark flavors and $B$ is the vacuum energy density, alternatively known as the bag constant. Here the individual quark flavor has energy density ${\rho}^f$ is related to $p^f$ as follows:
 \begin{equation}
p^f=\frac{1}{3}{{\rho}^f}. \label{eqeos2}
\end{equation}

 Now, the energy density of the deconfined quarks inside the bag is defined as
 \begin{equation}
{{\rho}}={\sum_{f=u,d,s}}{{\rho}^f}+B. \label{eqeos3}
\end{equation}

After substituting Eqs.~(\ref{eqeos2}) and (\ref{eqeos3}) into Eq.~(\ref{eqeos1}) one may derive the well known MIT bag model EOS for SQM given as
 \begin{eqnarray}\label{2.5}
 p_r=\frac{1}{3}\left(\rho-4B\right).
 \end{eqnarray}

In our numerical analysis, we consider the values of $B$ in the range $60-90~MeV/{{fm}^3}$. These values are consistent with the CERN data about quark-gluon plasma (QGP) and are compatible with the RHIC preliminary results~\citep{Heinz2000,Heinz2001,Blaizot2002,Burgio2002}. However, a relation of the information on the nuclear equation of state (EOS) due to high-energy heavy-ion collisions to physics of the interior of the ultra-dense compact stars, like neutron stars or strange quark stars, is unknown. In fact, there is a clear difference between the formation of QGP during the heavy-ion collision in the laboratory, and in the core of ultra-dense compact stars, needed to support the extreme inward gravitational pull. The possible QGP produced in the heavy-ion collision is characterized by high temperature and low baryon density, unlike the strange quark stars, made of QGP. In fact, it is by no means obvious that physics of strongly bound quark matter (particle physics) and physics of quark stars (astrophysics) are both mainly governed by strong interactions, because gravity is also the major player there for stars and cannot be ignored, e.g., vide following ref.~\citep{Lattimer2016}. In the original MIT bag model, the value of bag constant was $B=55~MeV/fm^3$. It was already verified~\citep{Farhi1984} that Witten's conjecture successfully describes the non-interacting and massless quarks with the $B$ values between $57-94~MeV/fm^3$. In support of our choice of $B$ we also mention the recent articles~ \citep{Burgio2002,Jaikumar2006,Bordbar2012,Kalam2013,Maharaj2014,Rahaman2014,Abbas2015,Arbanil2016,Moraes2016,Lugones2017,Alaverdyan2017} where the values of $B$ are also chosen in the range $60-90~MeV/fm^3$ for strange quark stars.

An analysis of a spherically symmetric stellar system demands the definition of the mass function as follows:
\begin{equation}\label{2.6}
m \left( r \right) =4\,\pi\int_{0}^{r}\!{{\rho}^{\it eff}} \left( r \right) {r}^{2}{dr}.
\end{equation}

Substitution of Eq.~(\ref{2.6}) into Eq.~(\ref{2.2}) yields 
\begin{eqnarray}\label{2.8}
 {{\rm e}^{-\lambda \left( r \right) }}=1-{\frac {2m}{r}},
 \end{eqnarray}
$m$ being the gravitational mass inside a sphere of radius $r$.
 
\begin{figure*}
\centering
    \subfloat{\includegraphics[width=4.5cm]{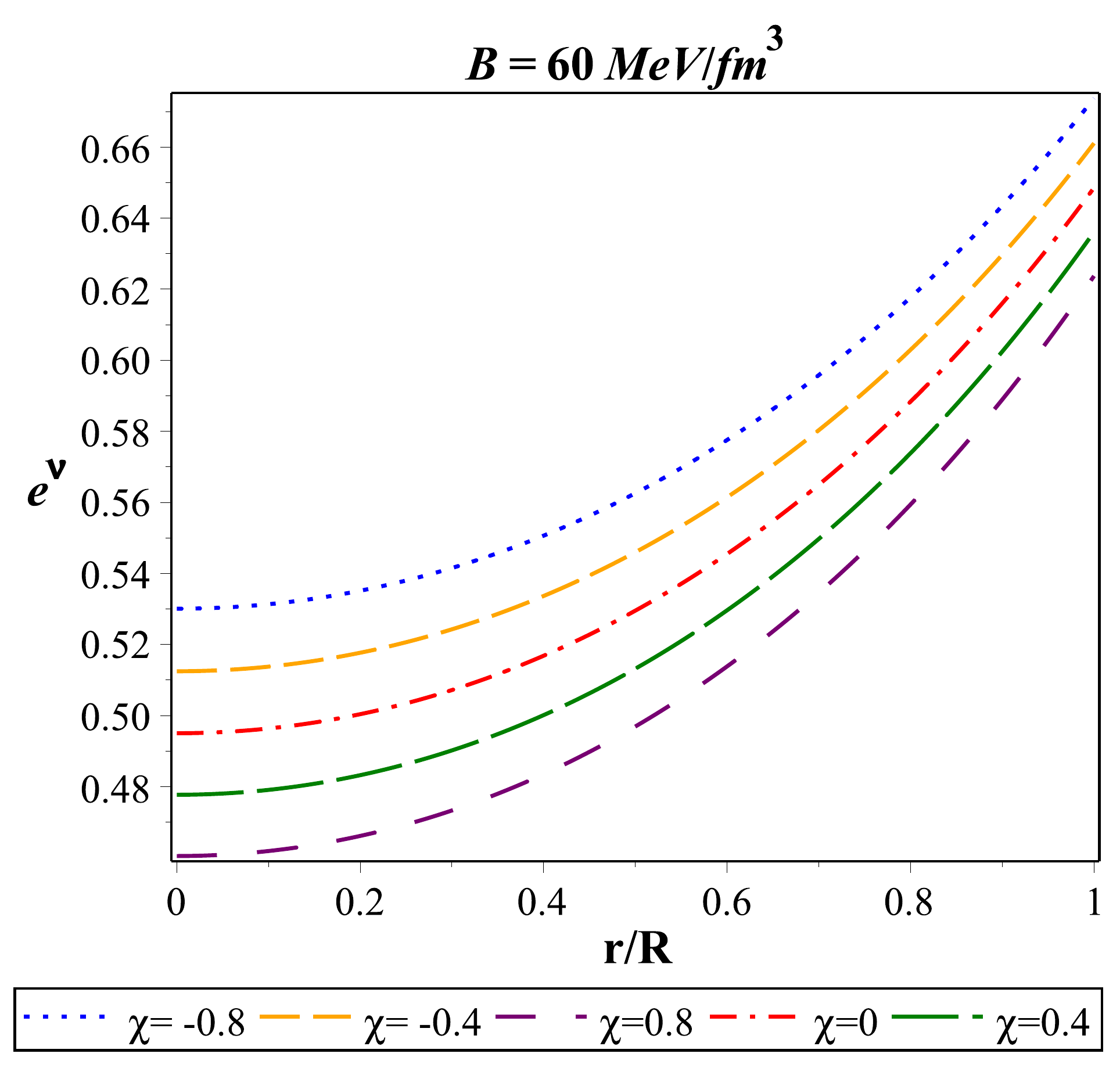}}
    \subfloat{\includegraphics[width=4.5cm]{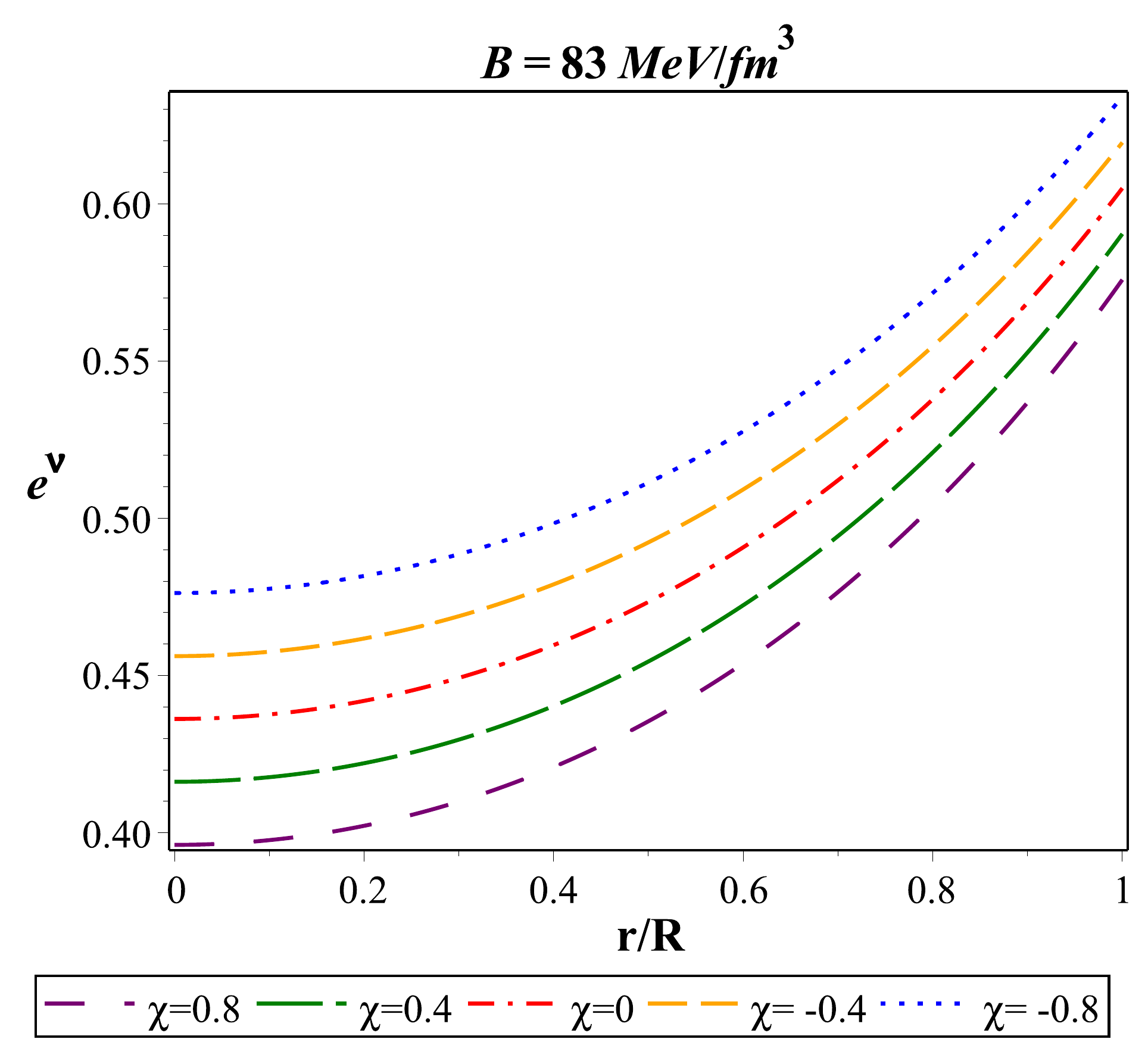}}
    \subfloat{\includegraphics[width=4.5cm]{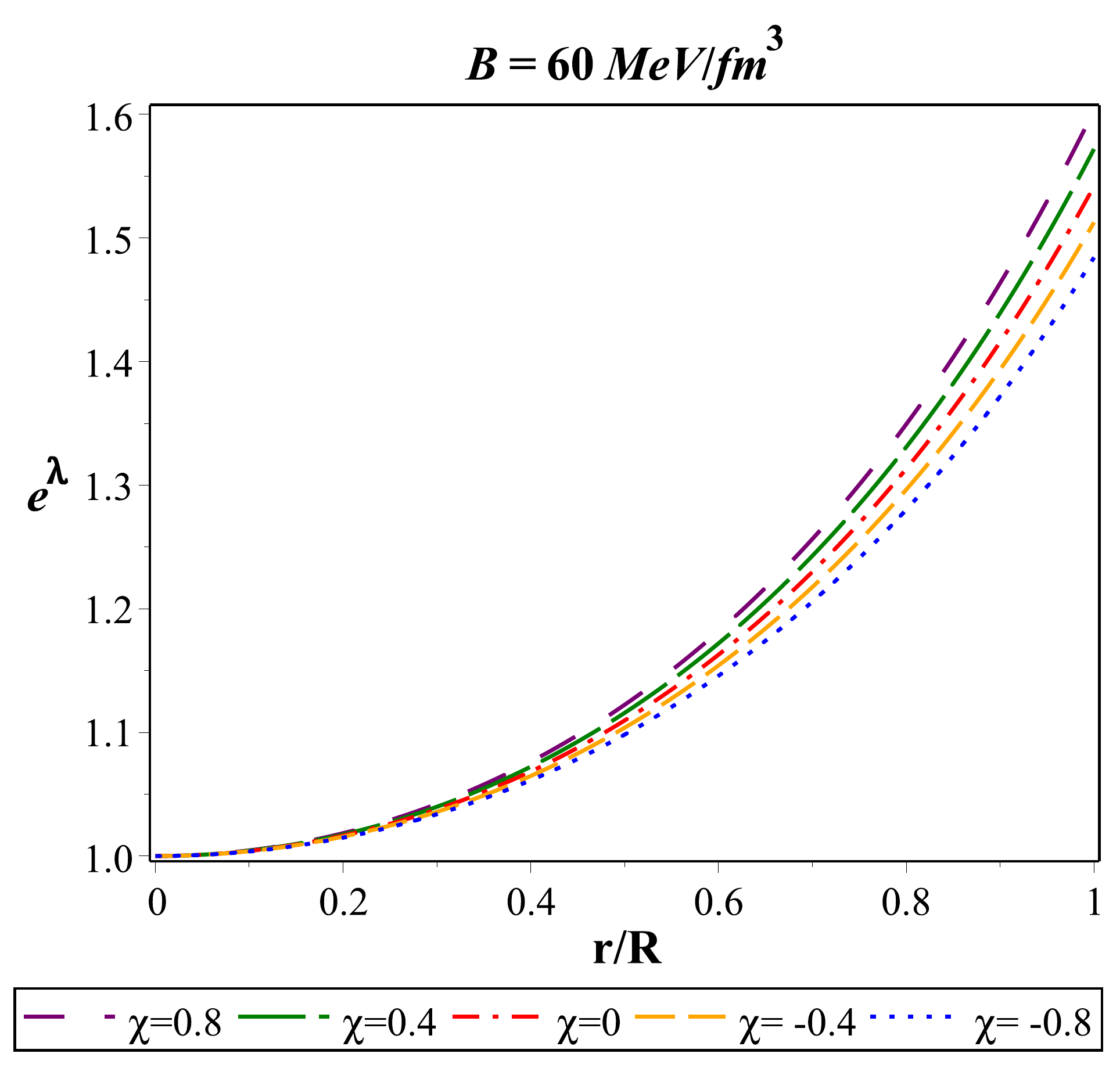}}
    \subfloat{\includegraphics[width=4.5cm]{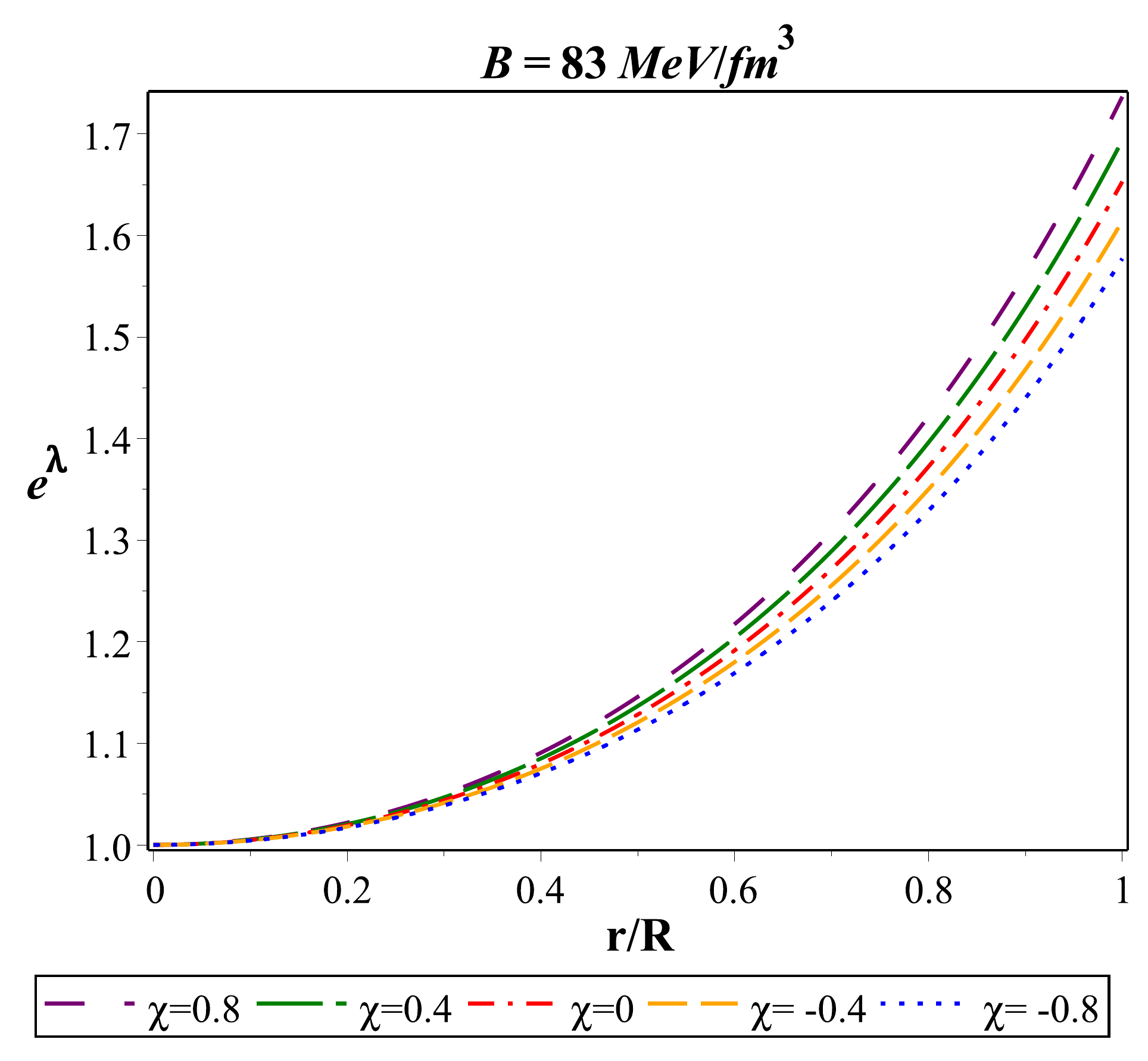}}
\caption{Variation of metric functions $e^{\nu}$ (first and second plots from the left) and $e^{\lambda}$ (third and fourth plots from the left) with the radial coordinate $r/R$ for $LMC\,X-4$.} \label{Fig1}
\end{figure*}

\section{Basic formalism of class 1 and general solutions} \label{sec3}
 Following~\citet{Lake2003}, in order to solve field Eqs.~(\ref{2.2})-(\ref{2.4}) we assume a specific form of the metric potential ${\nu}$ as
 \begin{equation}\label{3.1}
 {{\rm e}^{\nu}}=G \left( A{r}^{2}+1 \right) ^{n},
 \end{equation}
 where $A$, $G$ and $n$ are constants.
 
 For a physically valid solution the metric potential ${{\rm e}^{\nu}}$ must be free from the singularity and be a positive and monotonically increasing function of the radial coordinate. The metric potential we choose in Eq.~(\ref{3.1}) is consistent with all the above mentioned conditions and therefore is physically acceptable.
 
It is well known~\citep{Eisenhart1925} that space $V^{n+1}$ can be represented as an embedding class 1, i.e., $V^{n+1}$ can be projected as a hypersurface of a pseudo-Euclidean space $E^{n+2}$ if there exists a symmetric tensor $a_{mn}$  satisfying the following equations: \\
(i) The Gauss equation: 
\begin{eqnarray}\label{eqcls1.1}
R_{mnpq}=2e{a_{m\,[p}}{a_{q]n}},
\end{eqnarray}

\hspace{-0.7cm} (ii) The Codazzi equation:
\begin{eqnarray}\label{eqcls1.2}
a_{m\left[n;p\right]}-{\Gamma}^q_{\left[np\right]}a_{mq}+{{\Gamma}^q_{m}}\,{}_{[n}a_{p]q}=0,
\end{eqnarray}
where $e=\pm1$, $R_{mnpq}$ denotes curvature tensor and square brackets represent antisymmetrization. Here, $a_{mn}$ are the coefficients of the second differential form.~\citet{Eiesland1925}, after combining both the Gauss and Codazzi equations predicted a necessary and sufficient condition for the embedding class 1 in a more concise form as
 \begin{eqnarray}
R_{{0202}}R_{{1313}}=R_{{0101}}R_{{2323}}+R_{{1202}}R_{{1303}}.\label{3.2}
\end{eqnarray}

The required Riemannian symbols for the assumed interior spacetime Eq.~(\ref{2.1}) are 
\begin{eqnarray}\label{3.2.a}
 & R_{{0101}}=-\frac{1}{4}\,{{\rm e}^{\nu}} \left( -\nu^{{\prime}}\lambda^{{\prime}}+{\nu^{{\prime}}}^{2}+2
\,\nu^{{\prime\prime}} \right), \\ \label{3.2.b}
 & R_{{0202}}=-\frac{1}{2}\,r\nu^{{\prime}}{{\rm e}^{\nu-\lambda}},\\ \label{3.2.c}
 & R_{{1202}}=0,\\ \label{3.2.d}
 & R_{{1303}}=0,\\ \label{3.2.e}
 & R_{{1313}}=-\frac{1}{2}\,\lambda^{{\prime}}r \sin^2 \theta,\\ \label{3.2.f}
 & R_{{2323}}=-{r}^{2} {\sin^2 \theta} \left( 1-{{\rm e}^{-\lambda}} \right).
\end{eqnarray}

Substituting Eqs.~(\ref{3.2.a})-(\ref{3.2.f}) into Eq.~(\ref{3.2}) we find the differential equation
\begin{eqnarray}\label{3.3}
{{\rm e}^{\lambda}}{\lambda}^{{\prime}}{\nu}^{{\prime}}-{{\rm e}^{\lambda}}{{\nu}^{{\prime}
}}^{2}-2\,{{\rm e}^{\lambda}}{\nu}^{{\prime\prime}}+{{\nu}^{{\prime}}}^{2}+2{\nu}^{{\prime\prime}}=0.
\end{eqnarray}
 
 Now, substituting Eq.~(\ref{3.1}) into Eq.~(\ref{3.3}), we get the solution 
 \begin{equation}\label{3.4}
 {{\rm e}^{\lambda}}=\left[1-FA{r}^{2} \left( A{r}^{2}+1 \right) ^{n-2}\right], 
 \end{equation}
 where $F=4 CGA{n}^{2}$ and $C$ is the integration constant. 

We have shown variations of the metric potentials, viz., ${\rm e}^{\nu}$ and ${\rm e}^{\lambda}$, with respect to the radial coordinate in Fig.~\ref{Fig1}. It reveals that as the metric potentials are finite at the centre the system is free from the geometrical singularity.

\begin{figure}
\centering
    \subfloat{\includegraphics[width=4.5cm]{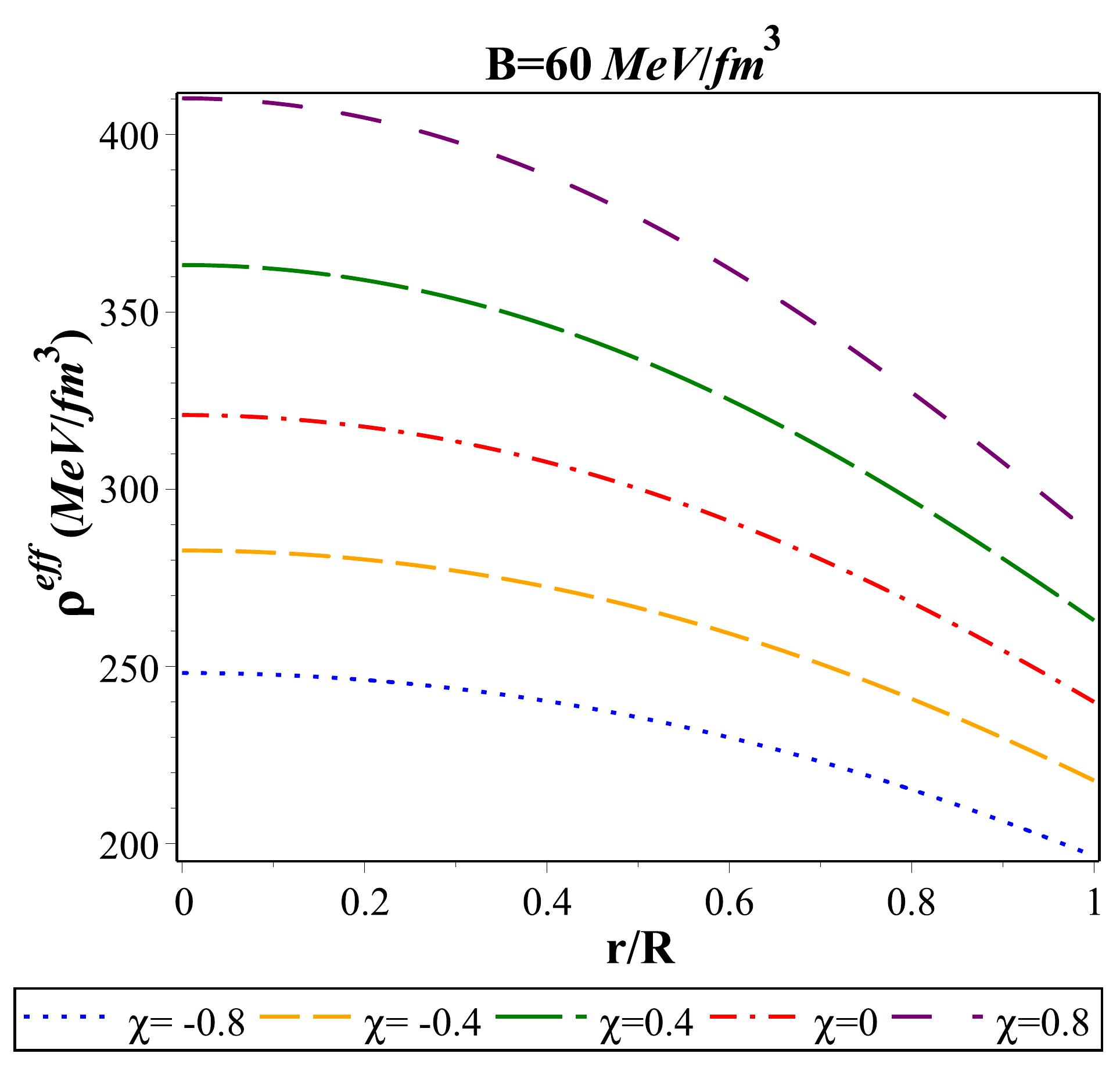}}
    \subfloat{\includegraphics[width=4.5cm]{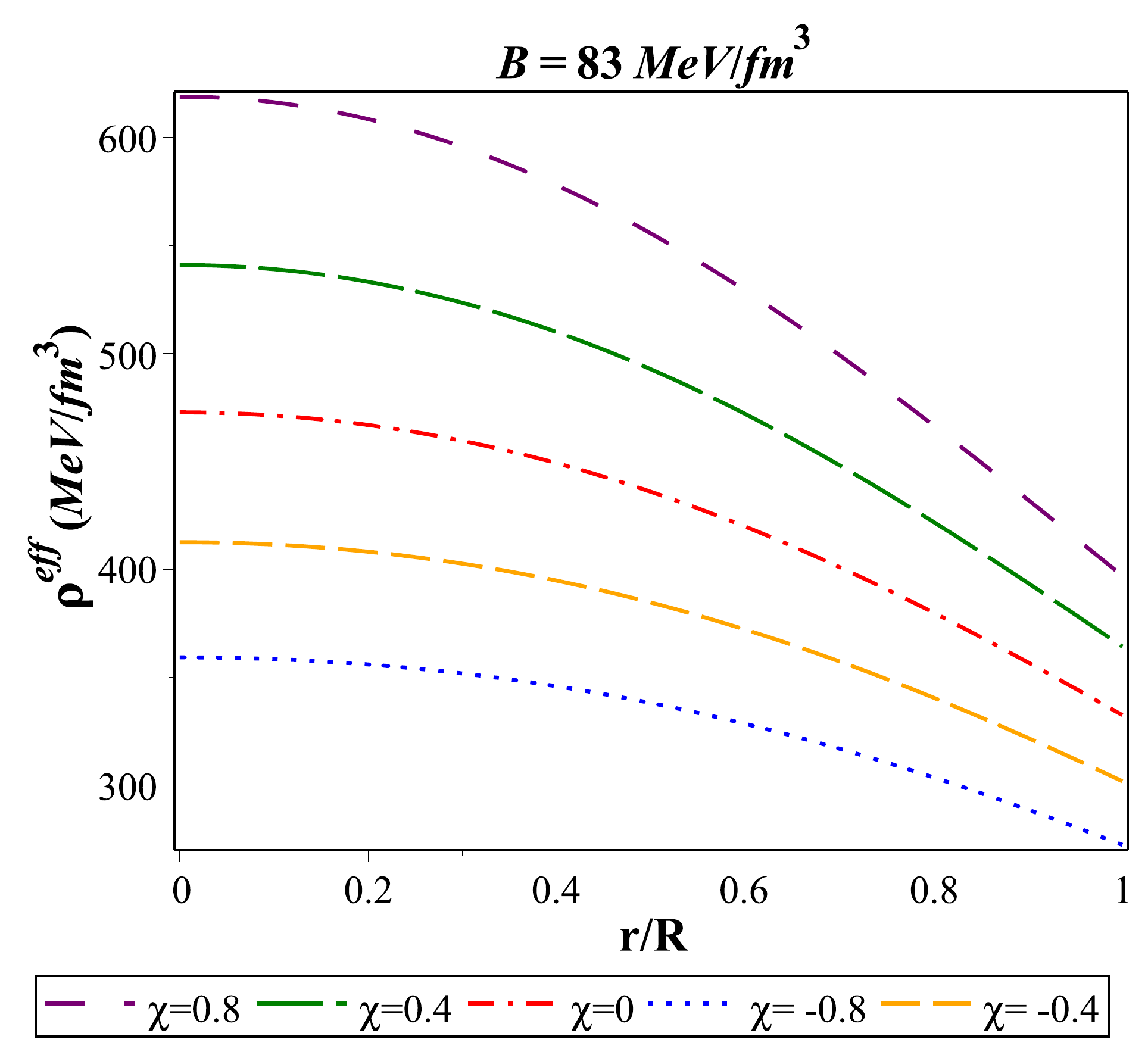}}
    \caption{Variation of matter density with the radial coordinate $r/R$ for $LMC\,X-4$} \label{Fig2}
\end{figure}
 
\begin{figure}
\centering
    \subfloat{\includegraphics[width=4.5cm]{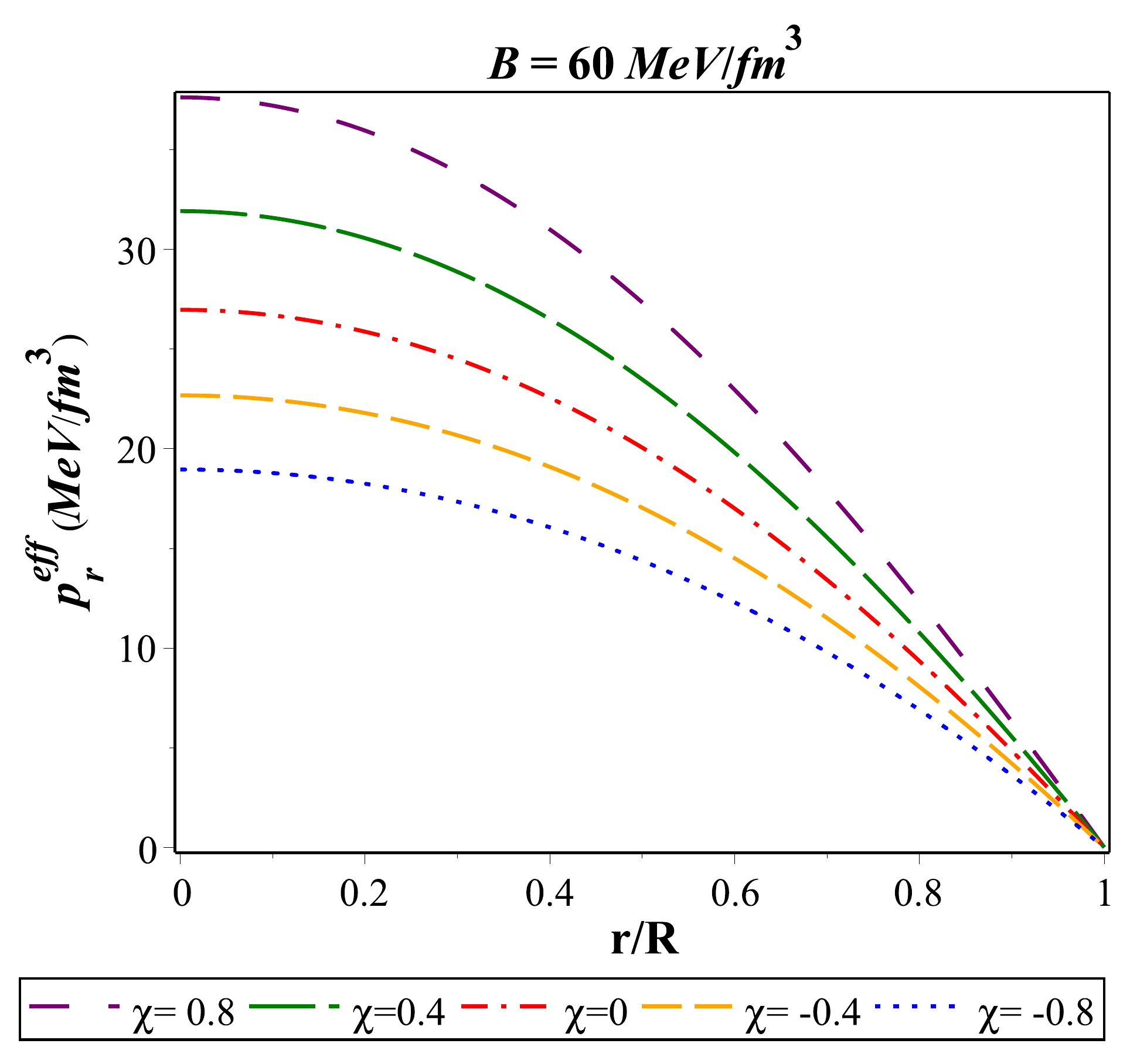}}
    \subfloat{\includegraphics[width=4.5cm]{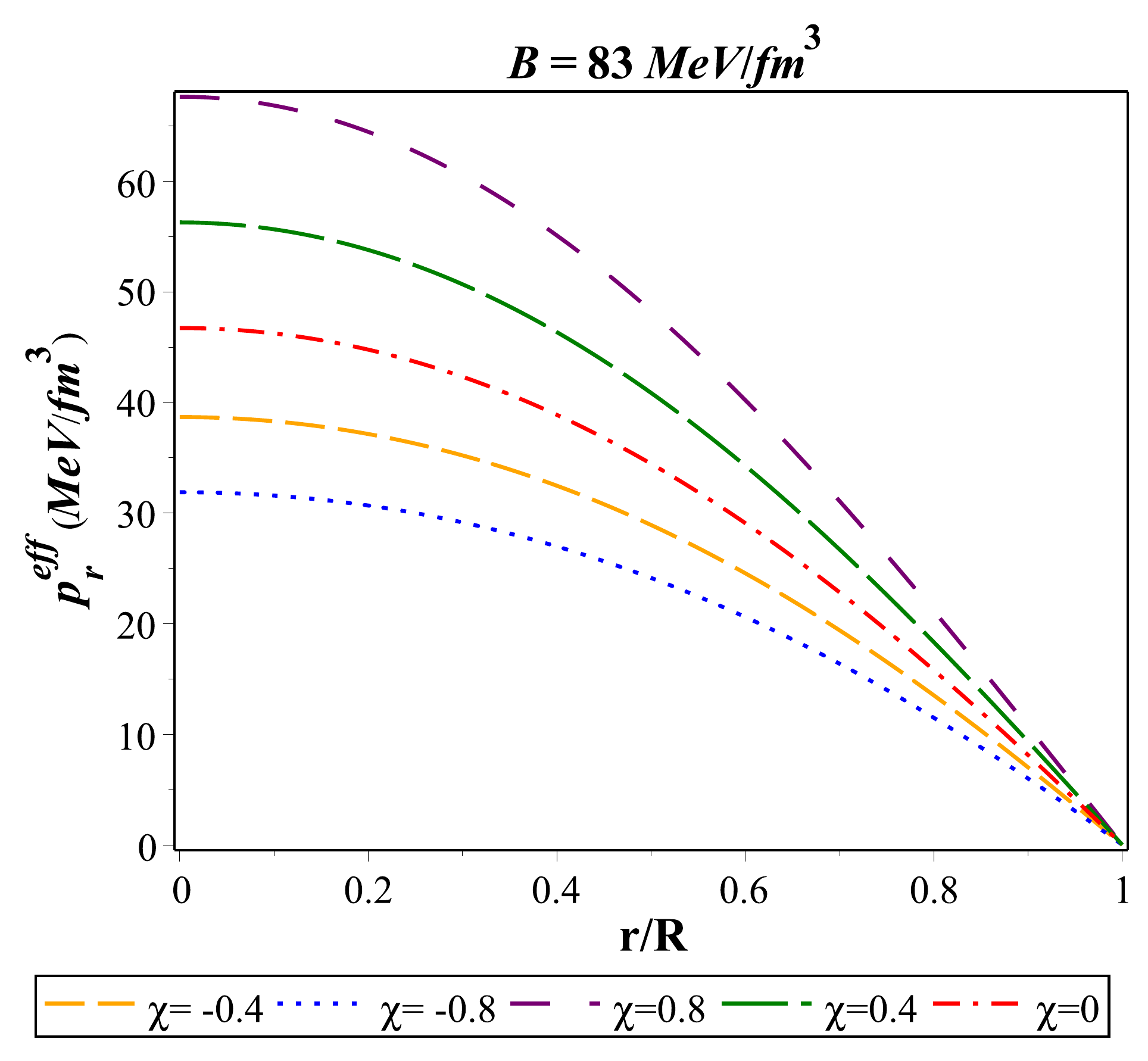}}
    \
    \subfloat{\includegraphics[width=4.5cm]{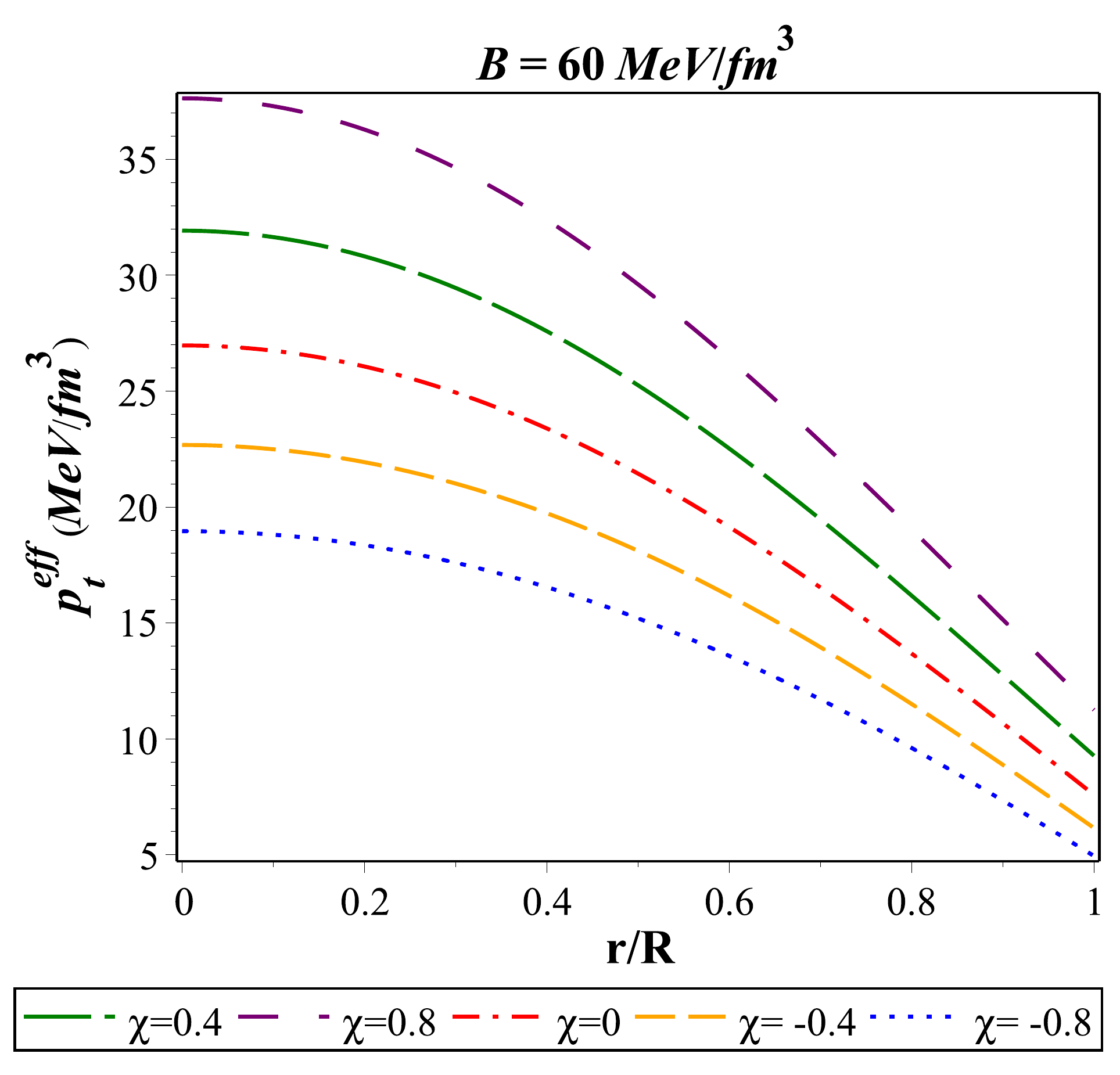}}
    \subfloat{\includegraphics[width=4.5cm]{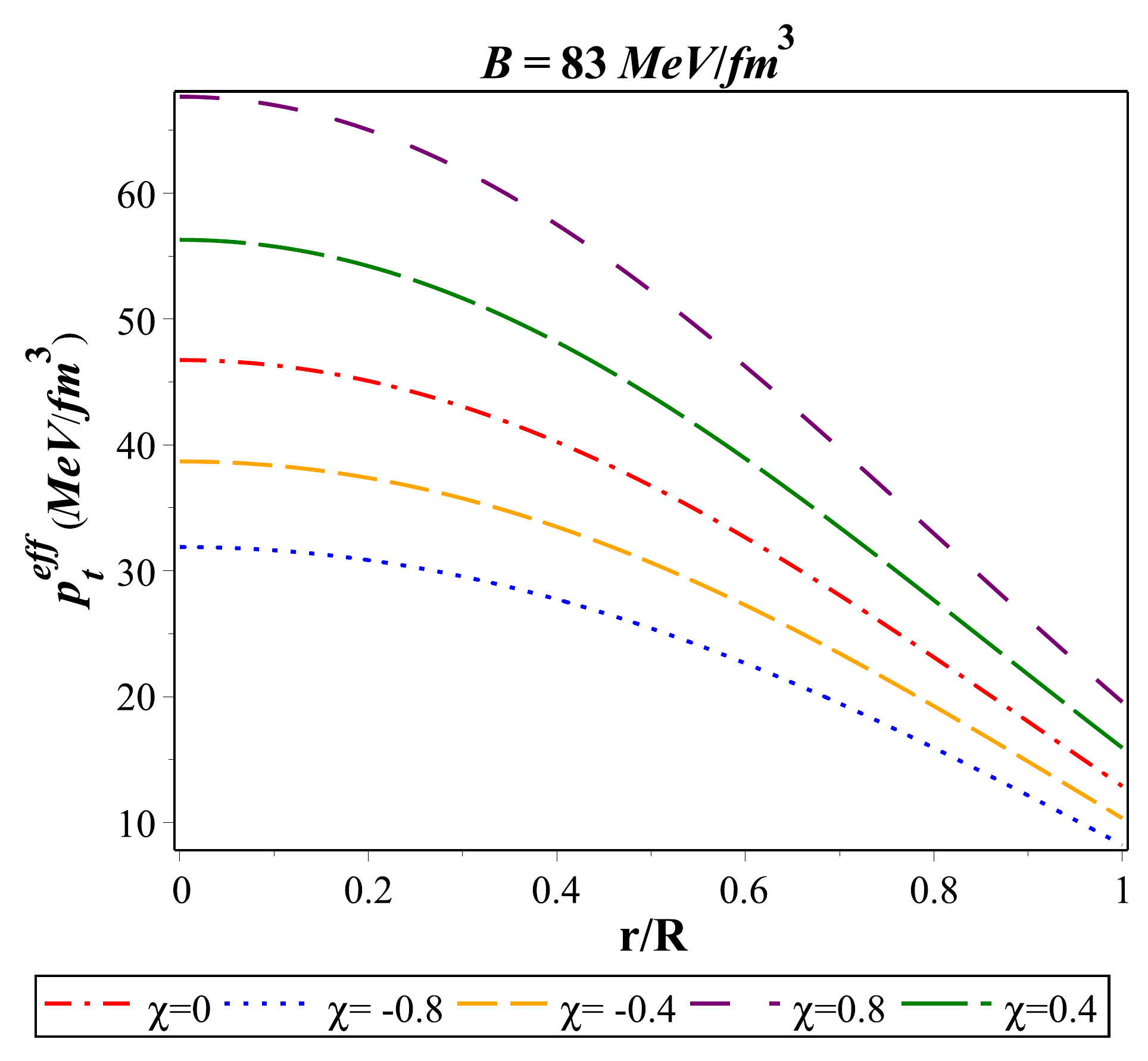}}
\caption{Variation of radial pressure $p^{\it eff}_r$ (upper panel) and tangential pressure $p^{\it eff}_t$ (lower panel) with the radial coordinate $r/R$ for $LMC\,X-4$.} \label{Fig3}
\end{figure}
 
 Substituting the metric potentials ${{\rm e}^{\nu}}$ of Eq.~(\ref{3.1}) and ${{\rm e}^{\lambda}}$ of Eq.~(\ref{3.4}) in Eqs.~(\ref{2.2})-(\ref{2.4}) and using the Eq.~(\ref{2.5}), we find the expression of the effective density~$\left({\rho}^{{\it eff}}\right)$, the effective radial pressure~$\left(p_r^{{\it eff}}\right)$ and the effective tangential pressure~$\left(p_t^{{\it eff}}\right)$ as
 \begin{eqnarray}\label{3.5}
&\qquad\hspace{-2cm} {\rho}^{{\it eff}}= {\frac {1}{24\pi \left(4\pi +\chi \right) {\rho_{{2}}}^{2}
 \left(2M{r}^{2}{\rho_{{2}}}^{n}-\rho_{{4}}{\rho_{{2}}}^{2} \right) ^{2}\rho_{{3}}}}\Big[384 \left( \rho_{{1}}{r}^{2}-{\frac {\chi}{96}} \right)\nonumber \\
&\qquad\hspace{-1cm} {r}^{2}{M}^{2}\rho_{{3}}{\rho_{{2}}}^{2+2n}+72\rho_{{4}} \lbrace([\frac{16}{3}\rho_{{1}}{r}^{2}+ ( {\frac {7n}{18}}-\frac{2}{9} ) \chi+\nonumber \\
&\qquad\hspace{-1cm}\pi ( n-\frac{1}{2} )] {r}^{2}M-\frac{16}{3} ( \rho_{{1}}{r}^{2}+{\frac{3\pi }{32}}+\frac{\chi}{32} ) \rho_{{3}} ) M{\rho_{{2}}}^{3+n}\nonumber \\
&\qquad\hspace{-1cm}-{\frac {\rho_{{4}}{\rho_{{2}}}^{4}}{72}} ([-96\rho_{{1}}{r}^{2}+ \big\lbrace(n-8) \chi-18\pi  \rbrace n ] {r}^{2}{M}^{2}+\nonumber \\
&\qquad\hspace{-2cm} \lbrace 192\rho_{{1}}{r}^{2}+18(\pi +\frac{\chi}{3}) n \rbrace M-96\rho_{{1}}\rho_{{3}} ) \rbrace \Big], 
\end{eqnarray}

\begin{eqnarray} \label{3.6}
&\qquad\hspace{-2cm} p_r^{{\it eff}}={\frac {\left( 2\,M{r}^{2}{\rho_{{2}}}^{n-2}-\rho_{{4}} \right) ^{-2}}{24\pi \left(4\pi +\chi \right) \left(2M{R}^{2}n-{R}^{3}n+M{R}^{2}-M{r}^{2} \right) ^{2} }}\nonumber \\
&\qquad\hspace{-1cm} \Big[-1536(\rho_{{1}}{r}^{2}-{\frac {\chi}{96}}) {r}^{2} (((n+\frac{1}{2}) {R}^{2}-\frac{1}{2}{r}^{2})M\nonumber \\
&\qquad\hspace{-1cm}-\frac{1}{2}{R}^{3}n)^{2}{M}^{2} ({\rho_{{2}}}^{n-2})^{2}+1536M(((\rho_{{1}}{r}^{2}-\frac{\pi}{32})\nonumber \\
&\qquad\hspace{-1cm}(n+\frac{1}{2}) {R}^{2}+\frac{1}{32}(-16\rho_{{1}}{r}^{2}+(\pi -\frac{\chi}{6}) n-\frac{\pi}{2}+\frac{\chi}{6} ) {r}^{2} ) M\nonumber \\
&\qquad\hspace{-1cm}-\frac{1}{2}(\rho_{{1}}{r}^{2}-\frac{\pi}{32}){R}^{3}n ) \rho_{{4}} (((n+\frac{1}{2} ) {R}^{2}-\frac{1}{2}{r}^{2} ) M\nonumber \\
&\qquad\hspace{-1cm}-\frac{1}{2}{R}^{3}n ) {\rho_{{2}}}^{n-2}-384{\rho_{{4}}}^{2} ((\rho_{{1}} (n+\frac{1}{2})^{2}{R}^{4}+\nonumber \\
&\qquad\hspace{-1cm}\frac{1}{32}(-32\rho_{{1}}{r}^{2}+\pi n)(n+\frac{1}{2}) {R}^{2}-{\frac {1}{384}} (-96\rho_{{1}}{r}^{2}+\nonumber \\
&\qquad\hspace{-1cm}n (\chi n+6\pi -2\chi )) {r}^{2}) {M}^{2}-{R}^{3} ((n+\frac{1}{2}) \rho_{{1}}{R}^{2}\nonumber \\
&\qquad\hspace{-1cm}-\frac{1}{2}\rho_{{1}}{r}^{2}+{\frac {\pi n}{64}} ) nM+\frac{1}{4}\rho_{{1}}{R}^{6}{n}^{2} )\Big], \\ \label{3.7}
&\qquad\hspace{-2cm} p_t^{{\it eff}}={\frac {1}{12{\rho_{{3}}}^{2}\pi (4\pi +\chi )(2M{r}^{2}{\rho_{{2}}}^{n-2}-\rho_{{4}}) ^{2}{\rho_{{2}}}^{2}}}\nonumber \\
&\qquad\hspace{-1cm}\Big[-192{\rho_{{2}}}^{2}{\rho_{{3}}}^{2}{r}^{2}{M}^{2}( \rho_{{1}}{r}^{2}-\frac{\pi}{8}-\frac{\chi}{24})({\rho_{{2}}}^{n-2})^{2}\nonumber \\
&\qquad\hspace{-1cm}-12\rho_{{2}}\rho_{{3}}\rho_{{4}}M ( ( 16\rho_{{1}}{r}^{2}+ ( n-\frac{3}{2} ) \pi +\frac{2}{3}\chi ( n-1 )) {r}^{2}M\nonumber \\
&\qquad\hspace{-1cm}-16 ( \rho_{{1}}{r}^{2}-\frac{\pi}{32}) \rho_{{3}} ) {\rho_{{2}}}^{n-2}+6 ((-8\rho_{{1}}{r}^{2}+ ((n-\frac{3}{2} ) \pi +\nonumber \\
&\qquad\hspace{-1cm} \frac{1}{3}( n-2 ) \chi) n) {r}^{2}{M}^{2}-\frac{1}{2}\rho_{{3}} (-32\rho_{{1}}{r}^{2}+\pi n ) M\nonumber \\
&\qquad\hspace{-1cm}-8\rho_{{1}}{\rho_{{3}}}^{2} ) {\rho_{{4}}}^{2}  \Big],
\end{eqnarray}  
where \\
${\rho_{{1}}\!=\! \left( \pi +\frac{\chi}{2} \right)\!\left( \pi +\frac{\chi}{4} \right) B}$, \\
${\rho_{{2}}=-{\frac {M{r}^{2}{R}^{-2} }{\left( 2Mn-Rn+M \right) }}+1}$, \\
${\rho_{{3}}={R}^{2} \left( 2Mn-Rn+M \right)}\\ $~and~\\
${\rho_{{4}}= \left( {\frac {n \left( -R+2M \right) }{2Mn-Rn+M}} \right)^{n-2} \left(-R+2M \right) {R}^{2}}$.

Variation of ${\rho}^{{\it eff}}$, $p_r^{{\it eff}}$ and $p_t^{{\it eff}}$ with the radial coordinate are shown in Figs.~\ref{Fig2} and~\ref{Fig3}. From these figures, we find that all three parameters are maximum at the centre, and they are decreasing monotonically to reach their minimum values at the surface, which confirms the physical validity of the predicted stellar model. These figures also feature that the density and pressure functions are finite at the centre, which confirms that our system is free from the physical singularity. 
  
The anisotropy of the system can be obtained by using Eqs.~(\ref{3.6}) and (\ref{3.7}) as 
 {\footnotesize{\begin{eqnarray}\label{3.8}
&\qquad\hspace{-7cm} \Delta=p_t^{{\it eff}} - p_r^{{\it eff}} \nonumber \\
&\qquad\hspace{-1cm}={\frac {1}{2\pi{\rho_{{3}}}^{2} \left( 4\pi +\chi \right) 
 \left( 2M{r}^{2}{\rho_{{2}}}^{n-2}-\rho_{{4}} \right) ^{2}{\rho_{{2}}}^{2}}}\Big[ -32 ( \rho_{{1}}{r}^{2}-\frac{\pi}{8}-\frac{\chi}{24} )\nonumber \\
&\qquad\hspace{-1cm} {M}^{2}{r}^{2}{\rho_{{2}}}^{2}{\rho_{{3}}}^{2} ( {\rho_{{2}}}^{n-2} )^{2}-2M (  ( 16\rho_{{1}}{r}^{2}+ 
( \pi +\frac{2}{3}\chi ) n-\frac{3}{2}\pi \nonumber \\
&\qquad\hspace{-1cm}-\frac{2}{3}\chi ) {r}^{2}M-16\rho_{{3}} ( \rho_{{1}}{r}^{2}-\frac{\pi}{32} )) \rho_{{2}}\rho_{{3}}\rho_{{4}}{\rho_{{2}}}^{n-2}+ ((-8\rho_{{1}}{r}^{2}+ \nonumber \\
&\qquad\hspace{-1cm}((\pi +\frac{\chi}{3} ) n-\frac{3}{2}\pi -\frac{2}{3}\chi ) n) {r}^{2}{M}^{2}-\frac{1}{2}\rho_{{3}} ( -32\rho_{{1}}{r}^{2}+n\pi ) M-\nonumber \\
&\qquad\hspace{-1cm}8\rho_{{1}}{\rho_{{3}}}^{2} ) {\rho_{{4}}}^{2} \Big]+\Big \lbrace 96( 2M{r}^{2}{\rho_{{2}}}^{n-2}+\rho_{{4}}) ^{2} ((( -2n-1 ) {R}^{2}\nonumber \\
&\qquad\hspace{-1cm}+{r}^{2}) M+{R}^{3}n)^{2}\pi ( \pi +\frac{\chi}{4}) \Big {\rbrace}^{-1} \Big{[} 384{M}^{2} ((( -2n-1 ) {R}^{2}+{r}^{2} ) M\nonumber \\
&\qquad\hspace{-1cm}+{R}^{3}n )^{2}{r}^{2} ( \rho_{{1}}{r}^{2}-{\frac {\chi}{96}} )( {\rho_{{2}}}^{n-2} )^{2}+384(( -2( \rho_{{1}}{r}^{2}-\nonumber \\
&\qquad\hspace{-1cm}\frac{\pi}{32} )( n+\frac{1}{2} ) {R}^{2}-\frac{1}{16}{r}^{2} (( \pi -\frac{\chi}{6} ) n-16\rho_{{1}}{r}^{2}-\frac{\pi}{2}+\nonumber \\
&\qquad\hspace{-1cm}\frac{\chi}{6} ) ) M+ ( \rho_{{1}}{r}^{2}-\frac{\pi}{32} ) {R}^{3}n ) M ((( -2n-1 ) {R}^{2}+{r}^{2} ) M+\nonumber \\
&\qquad\hspace{-1cm}{R}^{3}n ) \rho_{{4}}{\rho_{{2}}}^{n-2}+96{\rho_{{4}}}^{2} ( ( 4\rho_{{1}} ( n+\frac{1}{2} )^{2}{R}^{4}+\frac{1}{8}(-32\rho_{{1}}{r}^{2}+\nonumber \\
&\qquad\hspace{-1cm}n\pi ) ( n+\frac{1}{2} ) {R}^{2}-{\frac {{r}^{2}}{96}}( \chi {n}^{2}+ ( 6\pi -2\chi ) n-96\rho_{{1}}{r}^{2} )) {M}^{2}-\nonumber \\
&\qquad\hspace{-1cm}4{R}^{3} ( ( n+\frac{1}{2} ) \rho_{{1}}{R}^{2}-\frac{1}{2}\rho_{{1}}{r}^{2}+{\frac {n\pi }{64}} ) nM+\rho_{{1}}{R}^{6}{n}^{2} ) \Big{]}. 
 \end{eqnarray}}}

\begin{figure}
\centering
    \subfloat{\includegraphics[width=4.5cm]{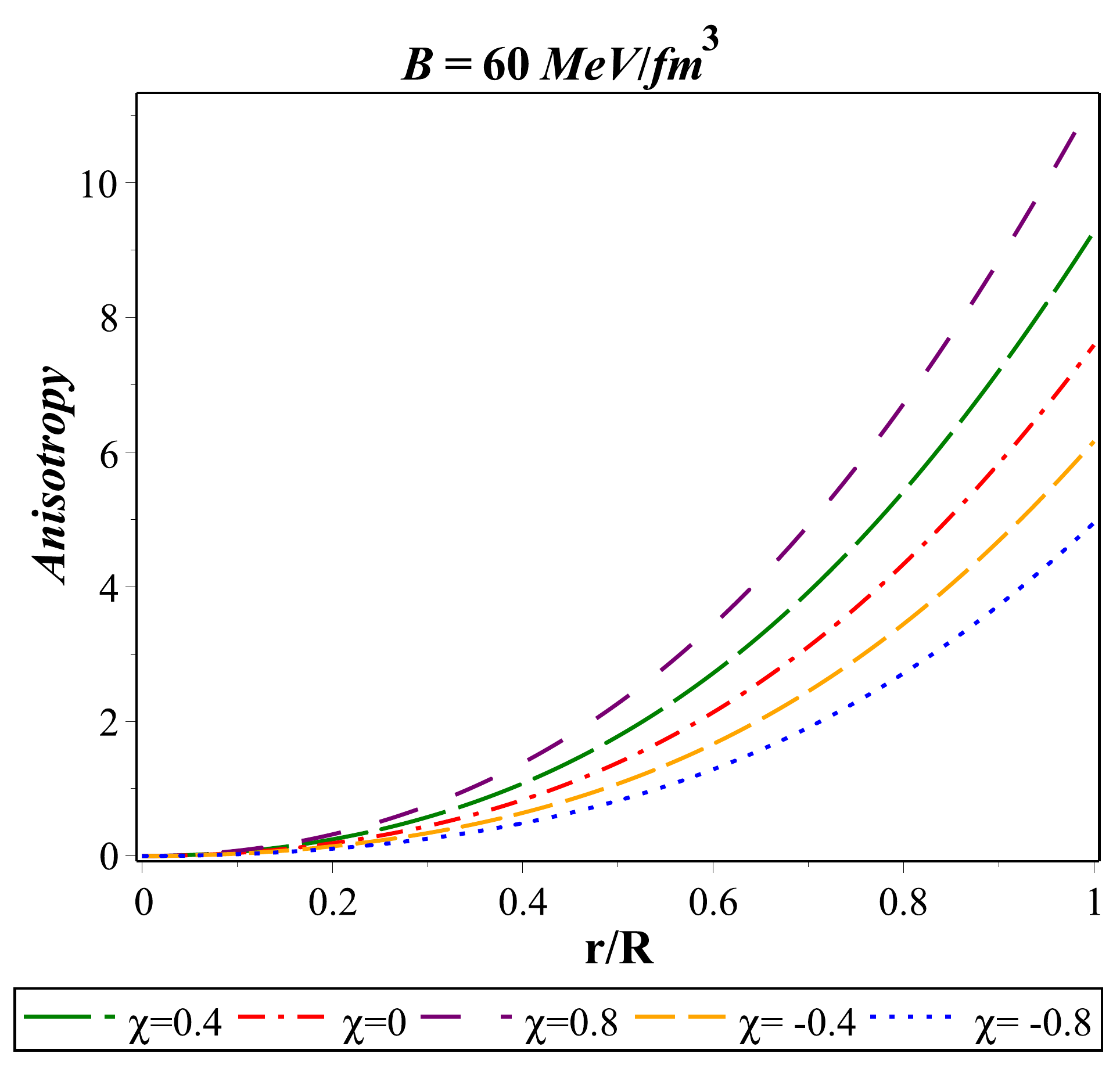}}
    \subfloat{\includegraphics[width=4.5cm]{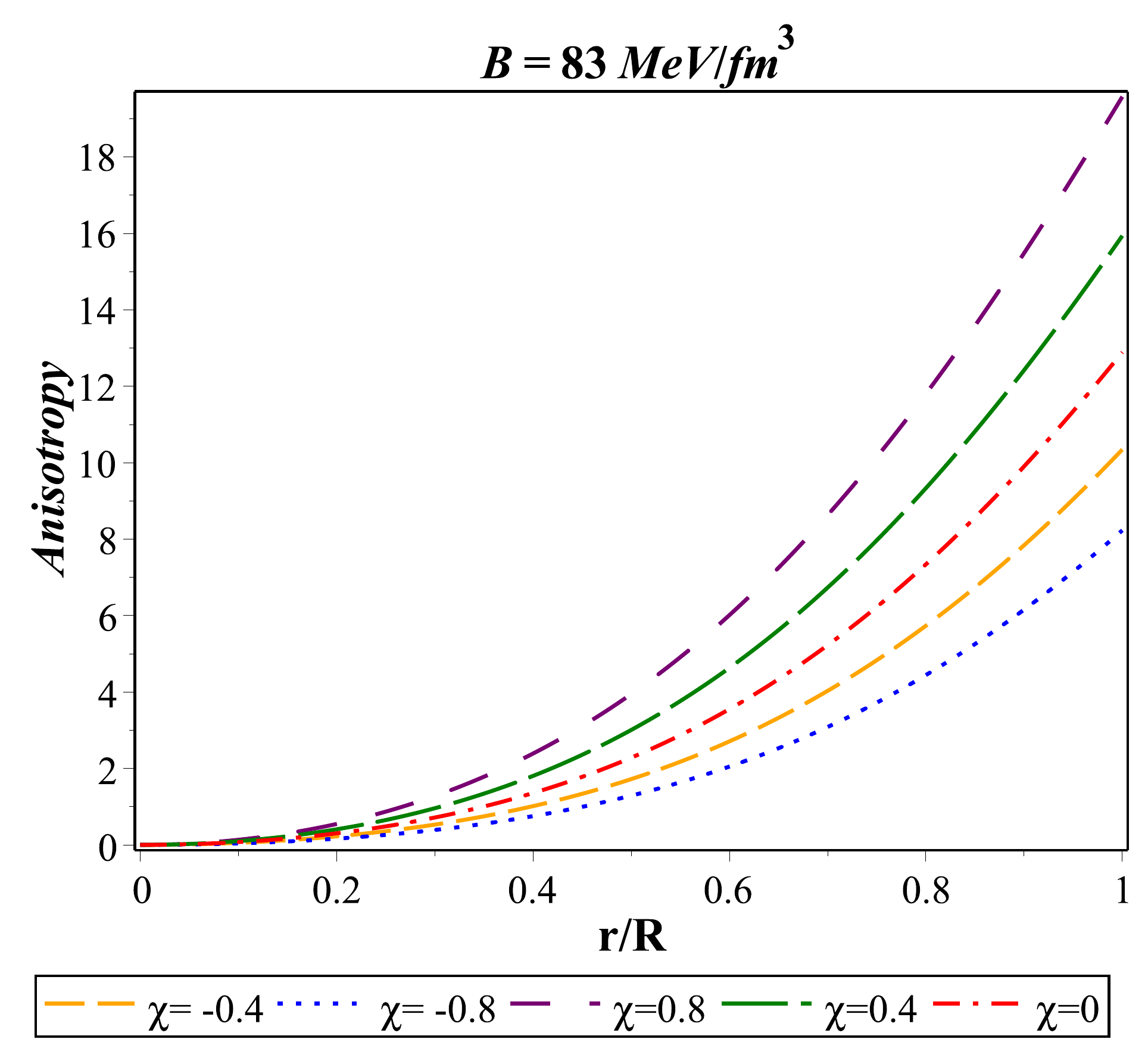}}
    \caption{Variation of anisotropy with the radial coordinate $r/R$ for $LMC\,X-4$.} \label{Fig4}
\end{figure}

  The behaviour of the anisotropy is shown in Fig.~\ref{Fig4}. In our study, we find in the case of anisotropic compact stellar system in the framework of modified $f\left(R,\mathcal{T}\right)$ gravity theory, that the anisotropy is zero at the centre and is maximum at the surface, which is matched well with the prediction of~\citet{Deb2017} in the case of Einsteinian gravity.

 \section{Matching boundary conditions and equllibrium configurations for the strange star candidates}\label{sec4}
 To derive the values of the model constants, we match our interior spacetime metric to the exterior Schwarzschild metric given by
 \begin{eqnarray}\label{4.1}
&\qquad\hspace{-0.5cm} ds^{2} =\left(1-\frac{2M}{r} \right) dt^{2} -\frac{dr^{2}}{\left(1-\frac{2M}{r}\right)}-r^{2} (d\theta ^{2}+\sin ^{2} \theta  d\phi ^{2} ).\nonumber \\
\end{eqnarray}

Here it is important to note that outside the stellar system ${\mathcal{T}=0}$ and hence, all the extra material terms
in the field equations due to the $f\left(R,\mathcal{T}\right)$ gravity theory are null so that the standard form of the Schwarzschild metric (\ref{4.1}) is still valid in the present case describing the exterior spacetime.
 
 The continuity of the metric potentials $g_{tt}$, $g_{rr}$ and the continuity of the second fundamental form, i.e., $\frac{d{g_{tt}}}{dr}$ at the surface $\left(r=R\right)$ yield the following equations:
 \begin{eqnarray}\label{4.2}
&\qquad G \left( A{R}^{2}+1 \right)^{n}=1-{\frac {2M}{R}},\\ \label{4.3}
&\qquad 1-FA{R}^{2} \left( A{R}^{2}+1 \right)^{n-2}= \left( 1-{\frac {2M}{R}} \right)^{-1},\\ \label{4.4}
&\qquad 2G \left( A{R}^{2}+1 \right)^{n-1}nAR={\frac {2M}{{R}^{2}}}.
 \end{eqnarray}
 
By solving Eqs.~(\ref{4.2})-(\ref{4.4}) we get the values of different constants as
 \begin{eqnarray}\label{4.5}
 &\qquad A=-{\frac {M}{{R}^{2} \left( 2Mn-Rn+M \right) }},\\ \label{4.6}
&\qquad G={\left(1-\frac {2M}{R}\right) \left[{\frac {n \left( 2M-R \right) }{2Mn-Rn+M}} \right] ^{-n}  },\\ \label{4.7}
&\qquad F={\frac {2\left(2Mn-Rn+M\right)}{R-2M}   \left[ {\frac {n \left( 2M-R
 \right) }{2Mn-Rn+M}} \right]^{2-n}  },\\ \label{4.8}
&\qquad C=-{\frac {{R}^{3}}{2M}},
 \end{eqnarray}
 where we already defined $F=4CGA{n}^{2}$. 
 
 For a physically acceptable stellar system, the radial pressure at the surface must be zero, i.e., ${p_r}\left(R\right)=0$ and it gives
 \begin{eqnarray}\label{4.8}
&\qquad\hspace{-1cm} n=-{\frac {2{M}^{2} \left( 12\pi -\chi \right) }{96B{\pi }^{2}{R
}^{4}+72B\pi {R}^{4}\chi+12B{R}^{4}{\chi}^{2}+12{M}^{2}\pi -{M
}^{2}\chi-18M\pi R}}. \nonumber \\
 \end{eqnarray}
 
In the framework of the modified $f\left(R,\mathcal{T} \right)$ gravity theory by using Eq.~(\ref{1.4}), the modified form of the energy-momentum tensor Eq.~(\ref{1.6}) can be written in a more explicit form as follows:
\begin{eqnarray}\label{4.9}
&\qquad\hspace{-4cm} -p_r^{{\prime}}-\frac{1}{2}\nu^{{\prime}} \left( \rho+p_r \right)+\frac{2}{r}\left({p_t}-{p_r}\right)\nonumber\\
&\qquad\hspace{1cm} +{\frac {\chi}{3(8\pi +2\chi)}}\left(3\rho^{{\prime}}-p_r^{{\prime}}-2p_t^{{\prime}} \right)=0.
\end{eqnarray} 

 Hence, the hydrostatic equilibrium equation for the anisotropic spherically symmetric compact stellar system in the framework of $f\left(R,\mathcal{T} \right)$ theory of gravity can be achieved using Eqs.~(\ref{2.3}), (\ref{2.8}), and (\ref{4.9}) as follows: 
\begin{eqnarray}\label{4.10}
&\qquad\hspace{-1cm} p_r^{\prime}=-\Big[(\rho+p_r)\lbrace 4\pi {p_r} {r^2}-\frac{\chi}{6}(3\rho-7{p_r}-2{p_t}){r^2}+\nonumber \\
&\qquad\hspace{-1cm} \frac{m}{r}+2(p_t-p_r)(1-\frac{2m}{r})\rbrace\Big] \Bigg/\Big[r\left(1-\frac{2m}{r}\right) \big \lbrace 1+\nonumber\\
&\qquad\hspace{1cm}\frac{\chi}{3(8\pi+2\chi)}\big(1-3\frac{d\rho}{d{p_r}}+2\frac{d{p_t}}{d{p_r}}\big) \big \rbrace\Big].
\end{eqnarray} 

To derive the exact values of radii of the strange star candidates for the different values of $n$, we have solved the hydrostatic equation (\ref{4.10}) by using Eqs. (\ref{2.5}) and (\ref{4.8}). To this end, we also consider the observed values for the mass of the stellar candidates and assume some particular values for $B$. Simultaneously, we derive the values of $\chi$ for the strange star candidates due to the parametric values of $n$ and $B$. For $\chi=0$ in Eq. (\ref{4.10}) we have the standard hydrostatic equation for the anisotropic spherically symmetric objects in the Einstein gravity.
 
\begin{figure*}
\centering
    \subfloat{\includegraphics[width=8cm]{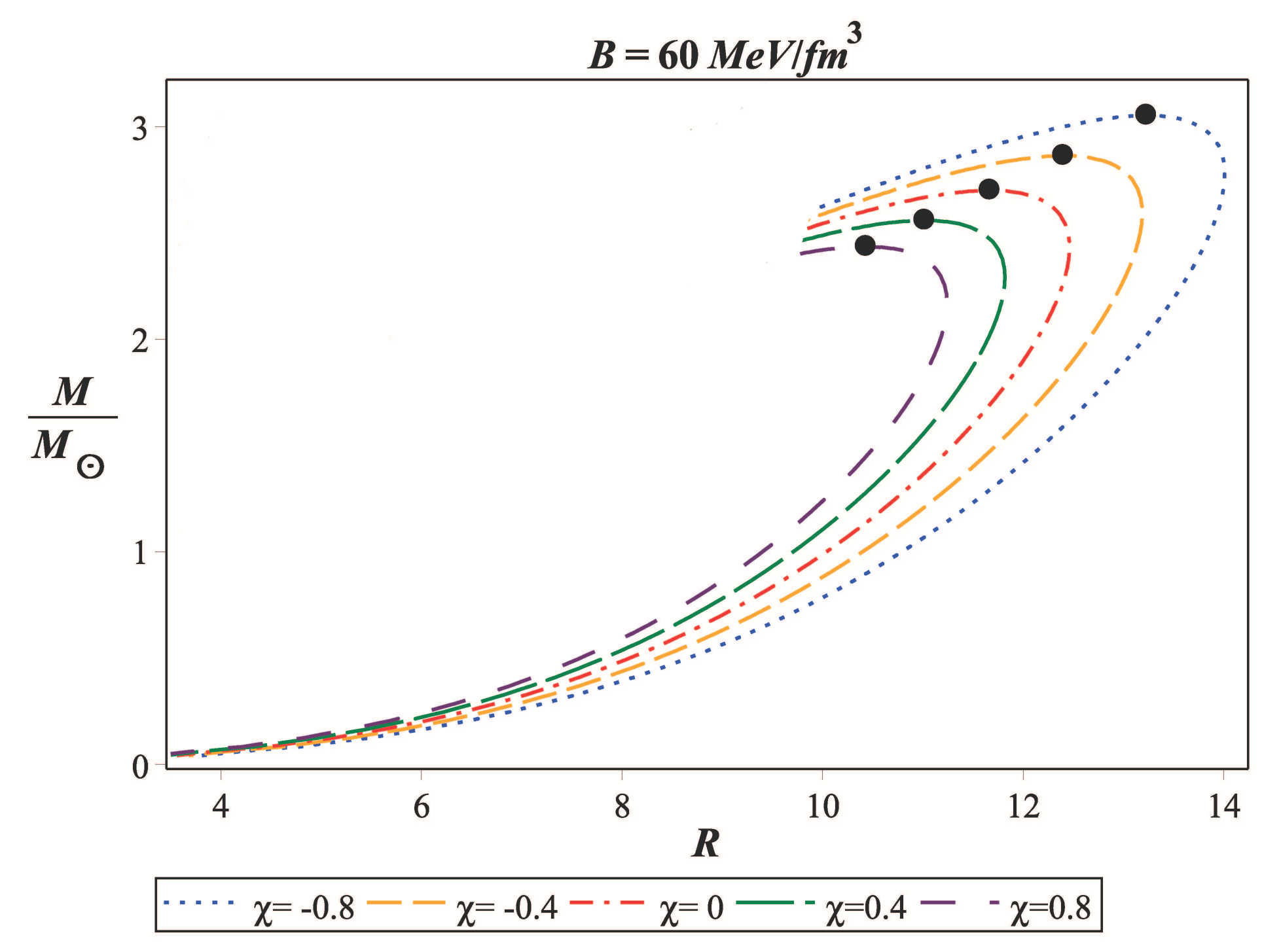}}  
    \subfloat{\includegraphics[width=8cm]{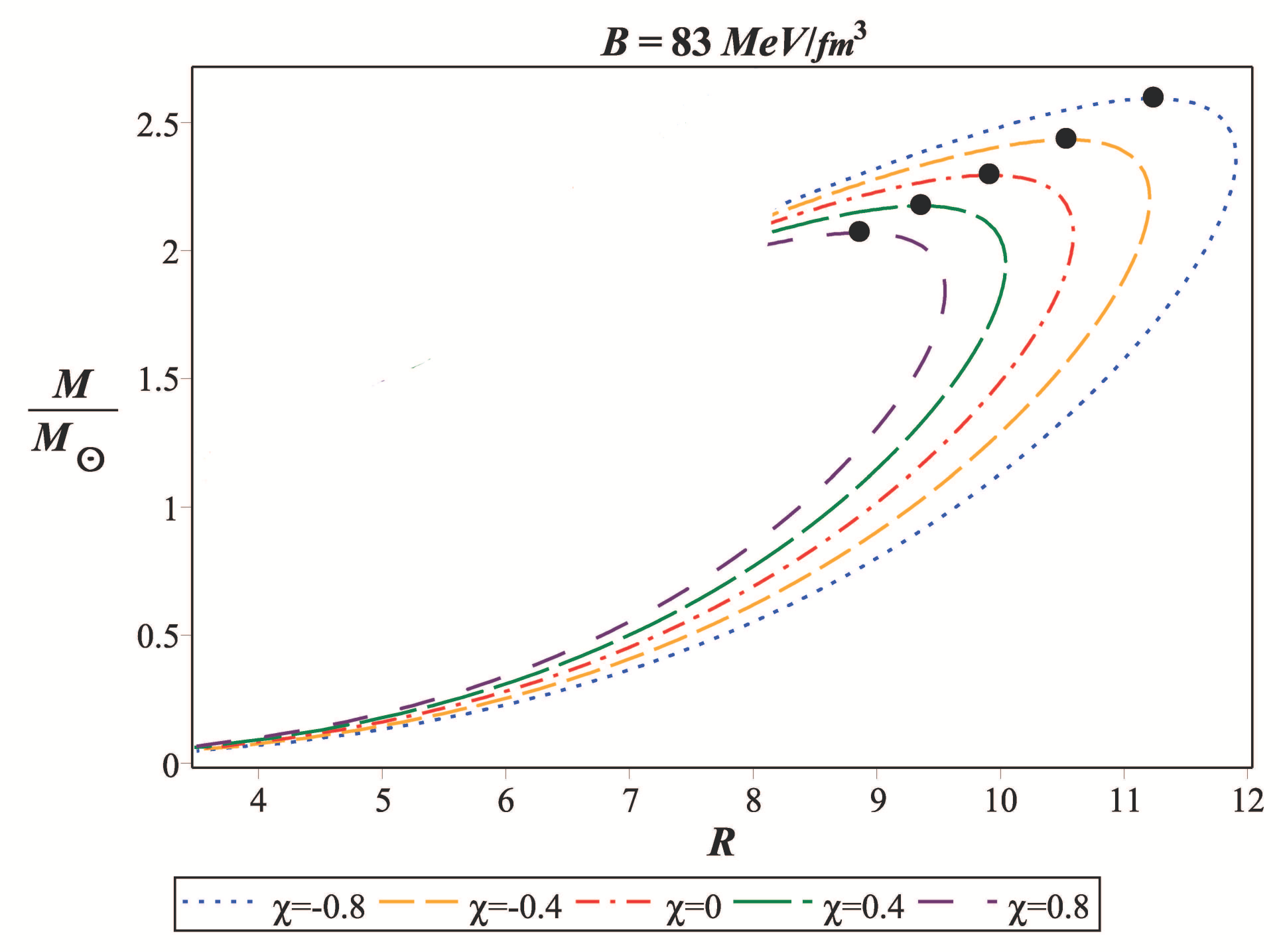}}
    \caption{Variation of the total mass normalized in the unit of Solar mass~$(M/{M}_{\odot})$ with the total radius $R$.} \label{FigMR}
\end{figure*}
 
In Fig.~\ref{FigMR} we present the behaviour of the total mass $(M)$, normalized in units of Solar mass~$(M_\odot)$, with respect to the total radius $(R)$ of the strange star candidates for the parametric values of $n$ and different chosen values of $B$. Interestingly, likewise the Einstein gravity, the mass-radius relation for the strange star candidates takes the usual form in the case of $f\left(R,\mathcal{T}\right)$ gravity theory too. We find that for $B=83~MeV/{fm}^3$, the maximal mass is $2.595~M_{\odot}$ when $\chi=-0.8$, and $2.070~M_{\odot}$ when $\chi=0.8$. The radii, corresponding to the maximum mass points, are $11.244~km$ for $\chi=-0.8$, and $8.867~km$ for $\chi=0.8$, whereas the obtained values of $n$ corresponding to the maximum mass points in the case $\chi=-0.8$, and $\chi=0.8$ are $n=2.210$ and $n=1.655$, respectively. Figure~\ref{FigMR} features that, as the values of $\chi$ decrease, the values of maximum mass points and the corresponding radii increase gradually. So, with the decreasing values of $\chi$, the stellar system turns into a more massive and less dense compact object.  


\begin{table*}
  \centering
    \caption{Physical parameters of the observed strange stars for $\chi=-0.4$ and $B=83 MeV/fm^3$.}\label{Table 1}
        \scalebox{0.9}{
\begin{tabular}{ ccccccccccccccccccccccccccc}
\hhline{=========}
Strange  & Observed & Predicted  & ${\rho}^{eff}_c$ &  ${\rho}^{eff}_0$ &  $p_c$ & $\frac{2M}{R}$ & $Z_s$ \\
Stars & Mass $(M_{\odot})$ & radius $(Km)$ &  $(gm/{cm}^3)$ & $(gm/{cm}^3)$ & $dyne/{cm}^2$ & & \\
\hline

PSR J1416-2230 & 1.97$\pm$0.04~\citep{Demorest2010} & 11.083 $\pm$ 0.037 & 9.799 $\times 10^{14}$ &	5.389 $\times 10^{14}$ &	1.385 $\times 10^{35}$ & 0.524 & 0.449\\

Vela X-1  & 1.77$\pm$0.08~\citep{Rawls2011} & 10.852 $\pm$ 0.108 &	8.756 $\times 10^{14}$ & 5.385 $\times 10^{14}$ &	1.058 $\times 10^{35}$ & 0.481 & 0.388\\

4U 1608-52 & 1.74$\pm$0.14~\citep{guver2010a}  & 10.811 $\pm$ 0.197 &	8.631 $\times 10^{14}$  &	5.385 $\times 10^{14} $ &	1.019 $\times 10^{35}$ & 0.475 & 0.380\\

PSR J1903+327 & 1.667 $\pm$0.021~\citep{Freire2011} &	10.703 $\pm$ 0.032 & 8.362 $\times 10^{14}$ & 5.384 $\times 10^{14}$ &	9.350 $\times 10^{34}$ & 0.459 & 0.360\\

4U 1820-30 &	1.58 $\pm$0.06~\citep{guver2010b}  & 10.562 $\pm$ 0.101 &	8.081 $\times 10^{14}$ & 5.383 $\times 10^{14}$ &	8.473 $\times 10^{34}$ & 0.441 & 0.337\\

Cen X-3   &	1.49 $\pm$0.08~\citep{Rawls2011}  & 10.403 $\pm$ 0.147 & 7.826 $\times 10^{14}$ & 5.382 $\times 10^{14}$ &	7.673 $\times 10^{34}$ & 0.422 & 0.315\\

EXO 1785-248 &	1.3 $\pm$0.2~\citep{ozel2009}  & 10.022 $\pm$ 0.435 & 7.375 $\times 10^{14}$ &	5.381 $\times 10^{14}$ & 6.264 $\times 10^{34}$ & 0.383 & 0.273\\

LMC X - 4 &	1.29 $\pm$0.05~\citep{Rawls2011} & 10.000 $\pm$ 0.109 &	7.355 $\times 10^{14}$ & 5.381 $\times 10^{14}$ &	6.201 $\times 10^{34}$ & 0.380 & 0.270\\

SMC X - 1 &	1.04 $\pm$0.09~\citep{Rawls2011} & 9.393 $\pm$ 0.242 &	6.885 $\times 10^{14}$ & 5.379 $\times10^{14}$ &	4.730 $\times 10^{34}$ & 0.327 & 0.219\\

SAX J1808.4-3658&	0.9 $\pm$0.3~\citep{Elebert2009} & 8.992 $\pm$ 0.934 &	6.663 $\times 10^{14}$ & 5.378 $\times 10^{14}$ &	4.036 $\times 10^{34}$ & 0..295 & 0.191\\

4U 1538-52 & 0.87$\pm$0.07~\citep{Rawls2011}  & 8.900 $\pm$ 0.2019 & 6.615 $\times 10^{15}$ & 5.378 $\times 10^{15}$ &	3.886 $\times 10^{34}$ & 0.288 & 0.185\\

Her X-1 & 0.85$\pm$0.15~\citep{Abubekerov2008} & 8.836 $\pm$ 0.481 & 6.589 $\times 10^{15}$  &	5.378 $\times 10^{15}$ & 3.804 $\times 10^{34}$ & 0.284 & 0.182\\

\hhline{=========} 
\end{tabular}  }
  \end{table*}



\begin{table*}
  \centering
    \caption{Derived values of constants due to the different strange star candidates for $\chi=-0.4$ and $B=83 MeV/fm^3$.}\label{Table 2}
       \scalebox{0.9}{
\begin{tabular}{ ccccccccccccccccccccccccccc}
\hhline{=========}
Strange Stars   & $n$ & $A$ & $C$ & $F$ & $G$   \\
\hline
PSR J1416-2230 & $8.706$ & $5.503 \times {10}^{-4}$ & $-234.252$  &	$-10.518$ & $0.269$\\

Vela X-1  & $16.181$   & $2.505 \times {10}^{-4}$ & $-244.756$ & $-20.814$ & $0.324$\\

4U 1608-52 & $18.311$   & $2.165 \times {10}^{-4}$ & $-246.165$ & $-23.759$ & $0.332$\\

PSR J1903+327 & $25.610$ &  $1.473 \times {10}^{-4}$ & $-249.321$ & $-33.929$ & $0.352$\\

4U 1820-30 & $44.938$ &  $7.948 \times {10}^{-5}$ & $-252.790$ & $-60.982$ & $0.376$\\

Cen X-3 & $161.229$  &  $2.101 \times {10}^{-5}$ & $-256.134$ & $-224.078$ & $0.400$\\

EXO 1785-248 & $-42.870$  & $-7.146 \times {10}^{-5}$ & $-262.481$ & $62.508$ & $0.453$\\

LMC~X-4 & $-40.731$ &  $-7.485 \times {10}^{-5}$ & $-262.778$ & $59.538$ & $0.456$\\

SMC~X-1 & $-17.373$  &  $-1.560 \times {10}^{-4}$ & $-270.120$ & $26.933$ & $0.529$\\

SAX J1808.4-3658 & $-13.676$ & $-1.866 \times {10}^{-4}$ & $-273.845$ & $21.882$ & $0.572$\\

4U~1538-52 & $-12.936$ & $-1.947\times {10}^{-4}$ & $-274.681$ & $20.834$ & $0.582$\\

Her X-1 & $-12.790$ & $-1.954 \times10^{-4}$ & $-275.122$ & $20.694$ & $0.588$\\

\hhline{=========}
\end{tabular}  }
  \end{table*}



\begin{table*}
  \centering
    \caption{Physical properties of $LMC~X-4$ due to different values of $\chi$ for $B=83 MeV/fm^3$.}\label{Table 3}
       \scalebox{0.9}{
\begin{tabular}{ ccccccccccccccccccccccccccc}
\hhline{=========}
Values  & Predicted     & obtained value & ${\rho}^{eff}_c$ &  ${\rho}^{eff}_0$ &  $p_c$ & $\frac{2M}{R}$ & $Z_s$  \\
of $\chi$  & radius $(Km)$ & of $n$      &  $(gm/{cm}^3)$   & $(gm/{cm}^3)$     & $dyne/{cm}^2$ & & \\ \hline

-0.8 & $10.398 \pm 0.117$ & $-17.317$ & $6.405 \times 10^{14}$ &	$4.855 \times 10^{14}$ & $5.112 \times 10^{34}$ & 0.366 & 0.256\\

-0.4 & $10.000 \pm 0.109$ & $-40.731$ & $7.355 \times 10^{14} $ &	$5.381 \times 10^{14}$ & $6.201 \times 10^{34}$ & 0.380 & 0.270\\

0  & $9.633 \pm 0.102$ & $106.881$ & $8.424 \times 10^{14}$ & $5.927 \times 10^{14}$ & $7.480 \times 10^{34}$ & 0.395 & 0.286\\

0.4 & $9.291 \pm 0.095$ & $21.980$ & $9.642 \times 10^{14}$ & $6.493 \times 10^{14}$ & $9.016 \times 10^{34}$ & 0.410 & 0.302\\

0.8 & $8.972 \pm 0.087$ & $12.152$ & $1.103 \times 10^{15}$ & $7.077 \times 10^{14}$ & $1.083 \times 10^{35}$ & 0.424 & 0.318\\

\hhline{=========} 
\end{tabular}  }
  \end{table*}



\begin{table*}
  \centering
    \caption{Physical properties of $LMC~X-4$ due to different values of $B$ for $\chi=0.4$.}\label{Table 4}
      \scalebox{0.9}{
\begin{tabular}{ ccccccccccccccccccccccccccc}
\hhline{=========}
Values  & Predicted  & Values of  & ${\rho}^{eff}_c$ &  ${\rho}^{eff}_0$ &  $p_c$ & $\frac{2M}{R}$ & $Z_s$  \\
of $B$ & radius $(Km)$& $n$ &  $(gm/{cm}^3)$ & $(gm/{cm}^3)$ & $dyne/{cm}^2$ & & \\
\hline
60  & $10.458 \pm 0.115$ & $106.707$ & $6.475 \times 10^{14}$ &  $4.689 \times 10^{14}$ & $5.113 \times 10^{34}$ & 0.364 & 0.254\\

80  & $9.423 \pm 0.097$ & $24.766$ & $9.188 \times 10^{14}$ &  $6.249 \times 10^{14}$ & $8.412 \times 10^{34}$ & 0.404 & 0.295\\

90  & $9.022 \pm 0.090$ & $17.807$  & $1.067 \times 10^{15}$ &  $7.030 \times 10^{14}$ & $1.043 \times 10^{35}$ & 0.422 & 0.315\\

\hhline{=========} 
\end{tabular}  }
  \end{table*}


 \section{Physical properties in $f(R,\mathcal{T})$ gravity}\label{sec5}
The physical properties of the stellar model in the $f\left(R,\mathcal{T}\right)$ are studied in the following subsections.
 
 \subsection{Energy conditions}\label{subsec5.1}
For our choice $f\left(R,\mathcal{T}\right)=R+2\chi\mathcal{T}$ the field equations can be summarized in Eq. (\ref{1.5}). The corresponding energy conditions, viz., the Weak Energy Condition (WEC), the Null Energy Condition (NEC), the Strong Energy Condition (SEC) and the Dominant Energy Condition (DEC) in $f\left(R,\mathcal{T}\right)$ gravity theory can be written down as~\citep{SC2013}
 {\small{\begin{eqnarray}\label{5.1.1}
&\qquad\hspace{-1.5cm} ~NEC:~~{{\rho}^{eff}}+{p^{eff}_r}\geq 0,~{{\rho}^{eff}}+{{p^{eff}_t}}\geq 0,\\ \label{5.1.2}
&\qquad\hspace{-1.5cm} ~WEC:~~{{\rho}^{eff}}+{{p^{eff}_r}}\geq 0,~{{\rho}^{eff}}\geq 0,~{{\rho}^{eff}}+{{p^{eff}_t}}\geq 0, \\ \label{5.1.3}
&\qquad\hspace{-1.5cm} ~SEC:~~{\rho}^{eff}+{p^{eff}_r} \geq 0,~{\rho}^{eff}+{p^{eff}_r}+2{p^{eff}_t} \geq 0, \\ \label{5.1.4}
&\qquad\hspace{-1cm}  ~DEC:~~{\rho}^{eff} \geq 0,~{\rho}^{eff}-{p^{eff}_r} \geq 0,~{\rho}^{eff}-{p^{eff}_t} \geq 0.
 \end{eqnarray}}}

\begin{figure}
\centering
    \subfloat{\includegraphics[width=4cm]{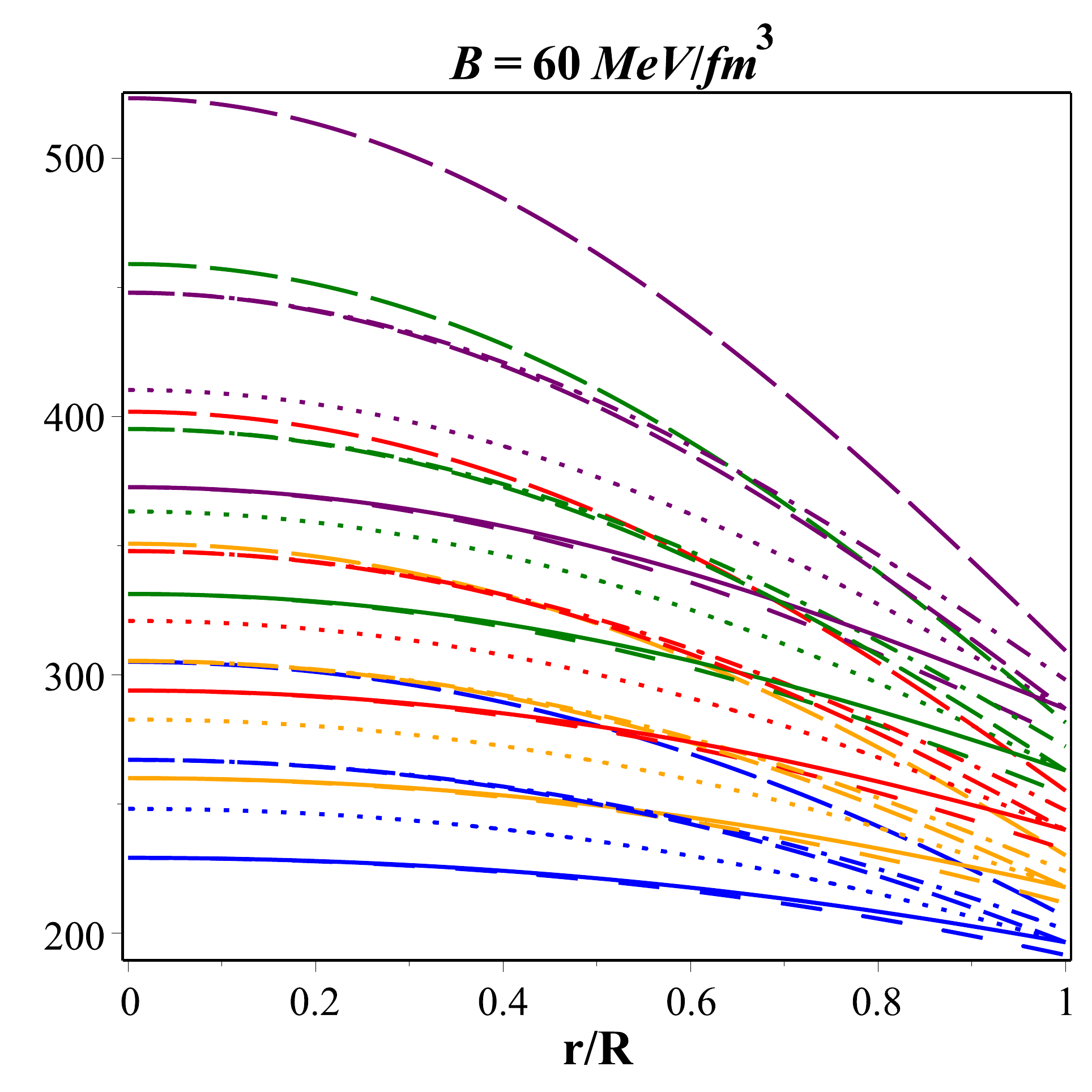}}
    \
    \subfloat{\includegraphics[width=4cm]{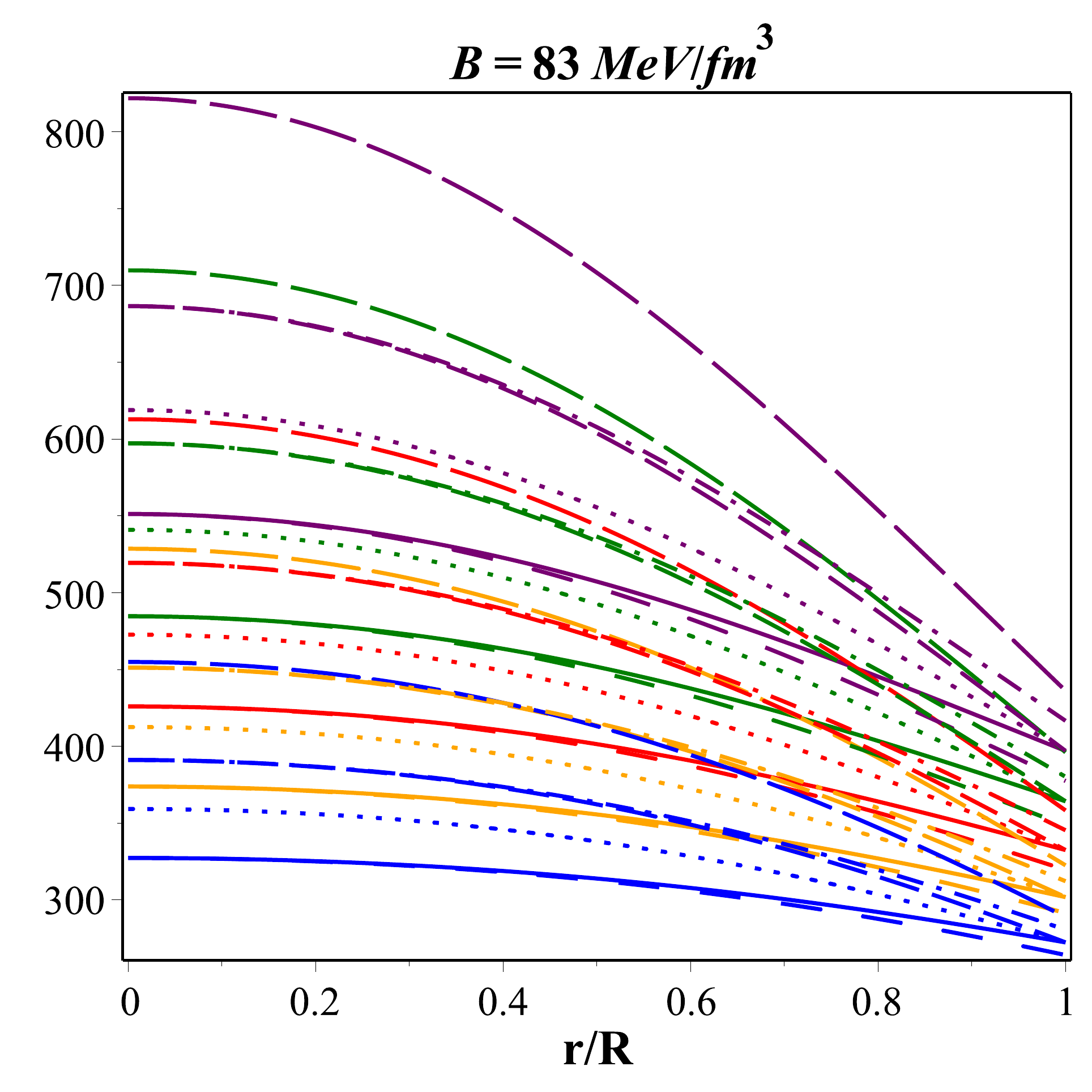}}
            \caption{Variation of energy conditions with the radial coordinate $r/R$ for $LMC\,X-4$. Here, dot, dash, dashdot, longdash, solid and spacedash linestyle represent respectively ${\rho}^{eff}$, ${\rho}^{eff}+p^{eff}_r$, ${\rho}^{eff}+p^{eff}_t$, ${\rho}^{eff}+p^{eff}_r+2\,p^{eff}_t$, ${\rho}^{eff}-p^{eff}_r$ and ${\rho}^{eff}-p^{eff}_t$, whereas purple, green, red, orange and blue color represent the cases with $\chi=0.8, 0.4, 0, -0.4$ and $-0.8$, respectively.} \label{Fig5}
\end{figure}
 
 For physical validity, the compact stellar system has to be consistent with the inequalities~(\ref{5.1.1})-(\ref{5.1.4}) simultaneously, in order to satisfy the energy conditions. Figure~\ref{Fig5} features that our system is well consistent with all the energy conditions for all chosen values of $B$ and $\chi$.

 \subsection{Stability of the system}\label{5.2}
 To study stability of the stellar model, we discuss (i) Modified TOV equation for $f\left(R,\mathcal{T}\right)$ gravity, (ii) Herrera cracking concept, and (iii) adiabatic Index.
 
 \subsubsection{Modified TOV equation for $f\left(R,\mathcal{T}\right)$ gravity} \label{subsubsec5.2.1}
In Eq.~(\ref{1.5}) we show the Einstein field equation for the $f(R,\mathcal{T})$ gravity theory which leads to the energy conservation of the effective energy-momentum tensor for our system as
 \begin{equation}\label{5.2.1.1}
 \nabla^{\mu}{T}^{eff}_{\mu\nu}=0.
 \end{equation}
 
We have already shown in Eq.~(\ref{4.9}) that the modified form of the generalized Tolman-Oppenheimer-Volkoff (TOV) equation for the $f\left(R,\mathcal{T}\right)$ gravity theory for our system is given by
 \begin{eqnarray*}\label{5.2.1.2}
&\qquad\hspace{-1cm} -p_r^{{\prime}}-\frac{1}{2}\nu^{{\prime}} \left( \rho+p_r \right)+\frac{2}{r}\left({p_t}-{p_r}\right) +{\frac {\chi \left(3\rho^{{\prime}}-p_r^{{\prime}}-2p_t^{{\prime}} \right)}{3(8\,\pi +2\,\chi)}}=0,\nonumber \\
 \end{eqnarray*}
where the first term is the gravitational force ($F_g$), the second term represents hydrodynamic force ($F_h$), the third term is the anisotropic force ($F_a$) and the last term represents force  ($F_m$) due to modification of the gravitational Lagrangian of the Einstein-Hilbert action. The modified TOV equation shows that the sum of the different forces in our system is zero, i.e, ${F_g}+{F_h}+{F_a}+{F_m}=0$.

\begin{figure}
\centering
    \subfloat{\includegraphics[width=4cm]{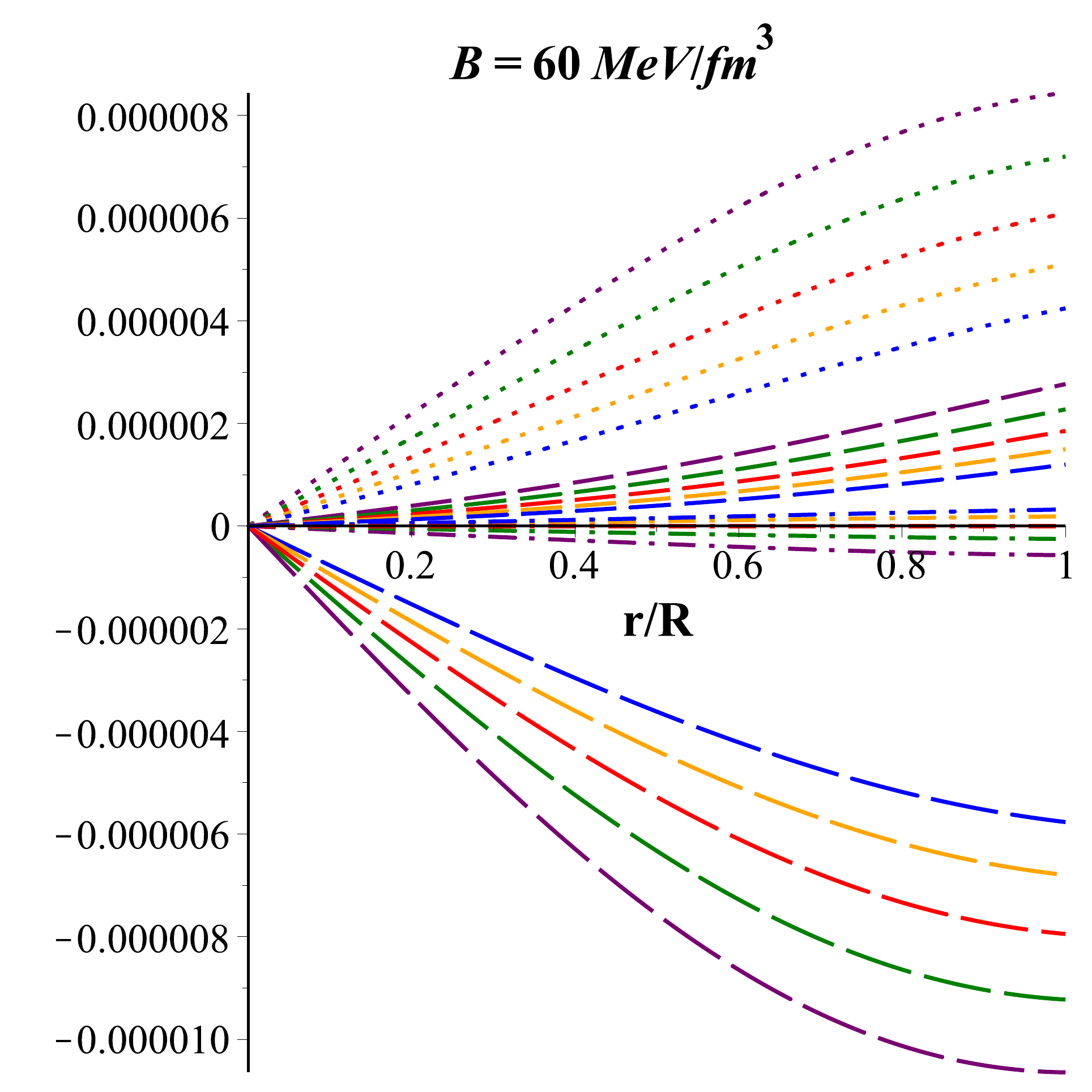}}
    \
    \subfloat{\includegraphics[width=4cm]{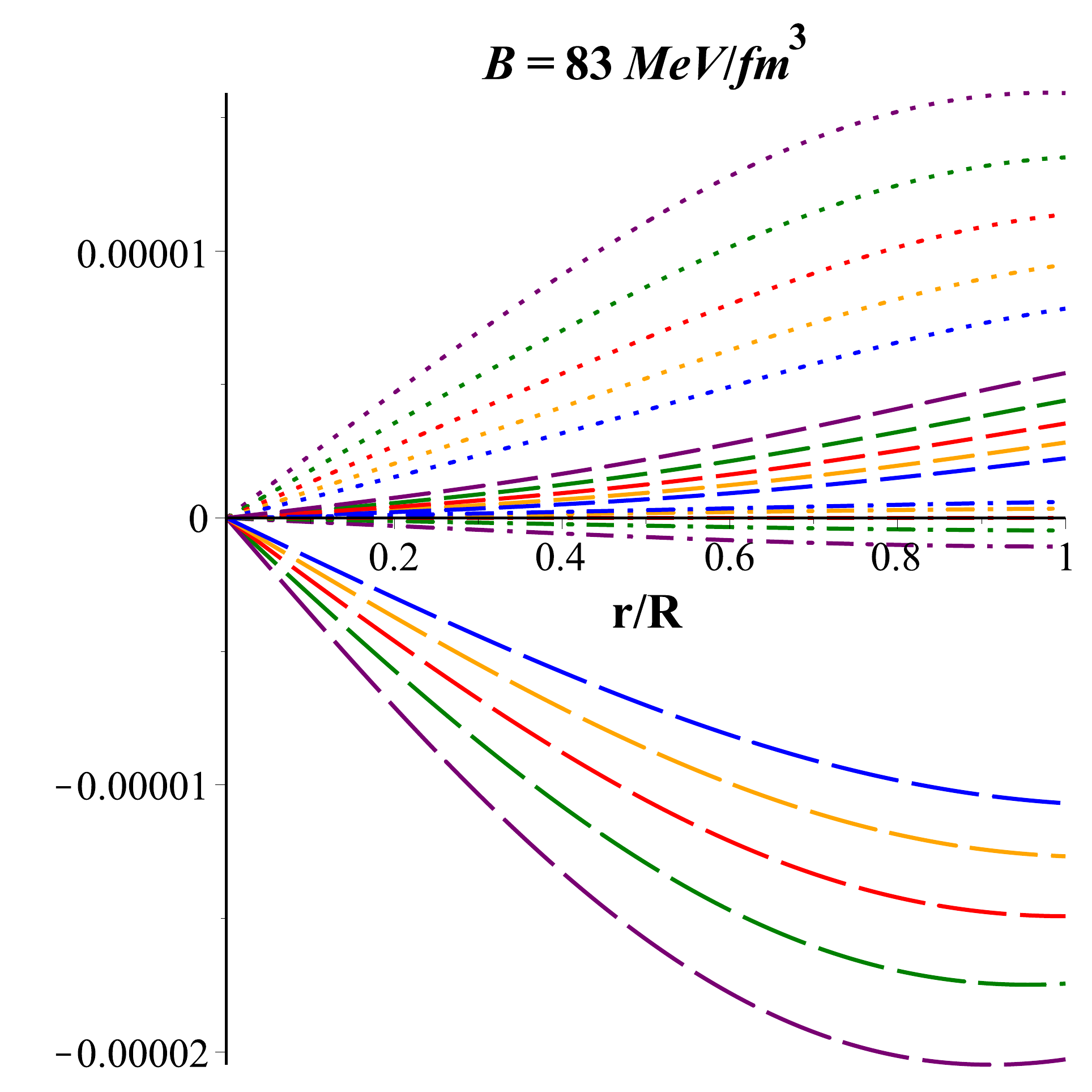}}
           \caption{Variation of forces with the radial coordinate $r/R$ for $LMC\,X-4$. Here, dot, dash, dashdot and longdash linestyle represent $F_h$, $F_a$, $F_m$ and $F_g$, respectively, whereas purple, green, red, orange and blue color represent the cases with $\chi=0.8, 0.4, 0, -0.4$ and $-0.8$, respectively.} \label{Fig6}
\end{figure}

 Fig.~\ref{Fig6} features that the equilibrium of the forces is achieved for our system, which ensures the stability of the proposed stellar system. This figure shows that in the case of $\chi>0$ the sum of the gravitational force and the force arising due to modified $f\left(R,\mathcal{T}\right)$ gravity theory are counterbalanced by the combined effect of the hydrodynamic force and anisotropic force, whereas in the case of $\chi<0$ the resultant effect of $F_h$,~$F_a$ and $F_m$ counterbalances the inward pull due to $F_g$. Hence, for $\chi<0$ the effect of $F_m$ acts along the outward direction and shows repulsive nature, whereas for $\chi>0$ we find $F_m$ acts along the inward direction and behaves like an attractive force.

\subsubsection{Herrera cracking concept}\label{subsubsec5.2.2}
For a stable stellar configuration, square of the radial~$\left(v^2_{r}\right)$ and tangential~$\left(v^2_{t}\right)$ sound speeds should lie within limit [0,1], which is known as the condition of causality~\citep{Herrera1992}. According to the Herrera cracking concept~\citep{Herrera1992,Abreu2007} for a potentially stable region, $v^2_{r}$ should be greater than $v^2_{t}$ and the difference of square of the sound speeds should maintain same sign throughout the matter distribution, i.e., no cracking. So, causality and Herrera cracking concept imply:~$i)~0<v^2_{r}<1$~and~$0<v^2_{t}<1$,~$ii)~0<\mid v^2_{t} - v^2_{r}\mid <1$. For our system, squares of the sound speeds are defined by
\begin{eqnarray}\label{5.2.2.1}
&\qquad v^2_{r}=\frac{d {p^{\it eff}_r}}{d {\rho}^{\it eff}},\\\label{5.2.2.2}
&\qquad v^2_{t}=\frac{d {p^{\it eff}_t}}{d {\rho}^{\it eff}}.
\end{eqnarray}

\begin{figure}
\centering
    \subfloat{\includegraphics[width=4.5cm]{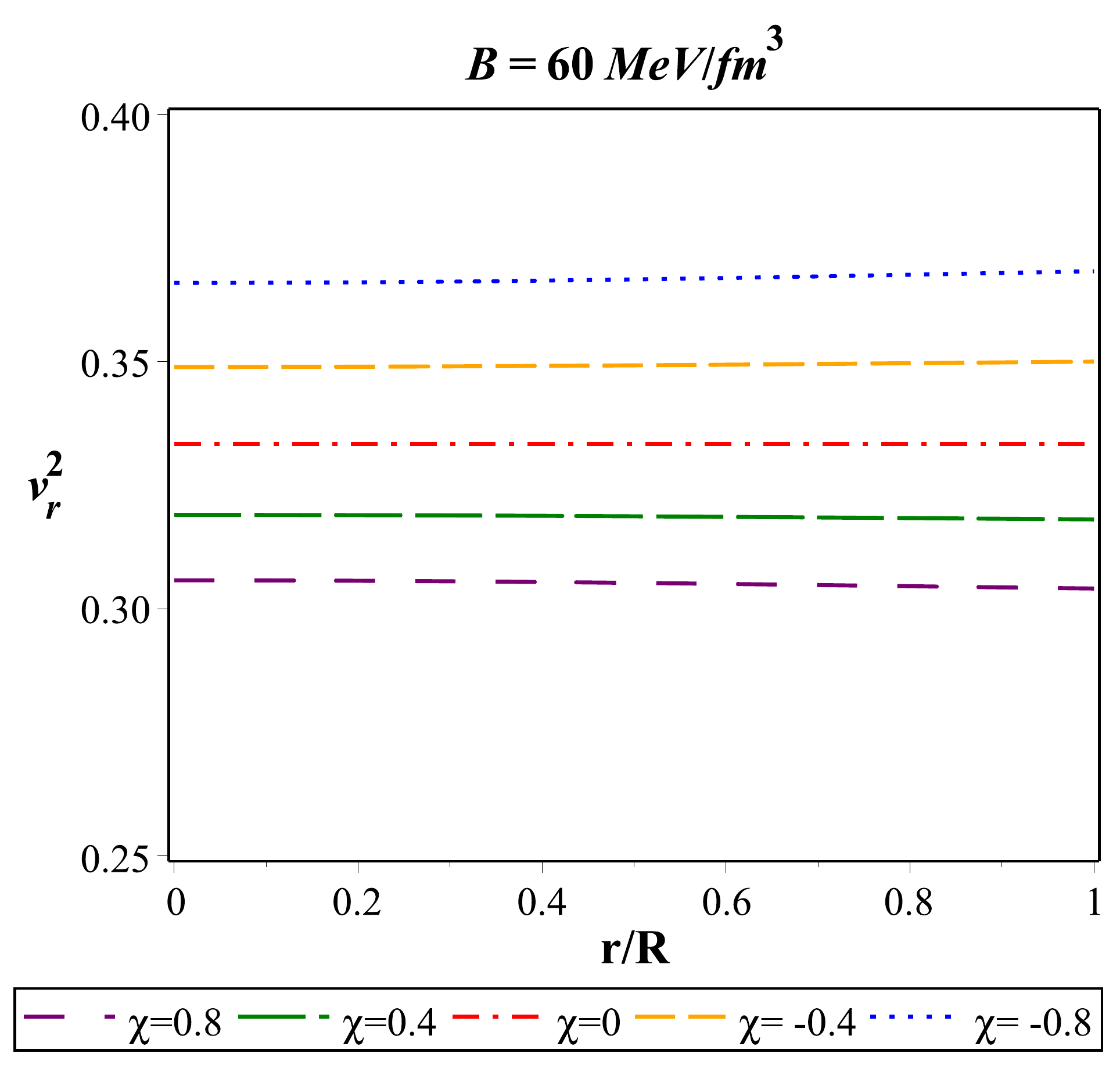}}
    \subfloat{\includegraphics[width=4.5cm]{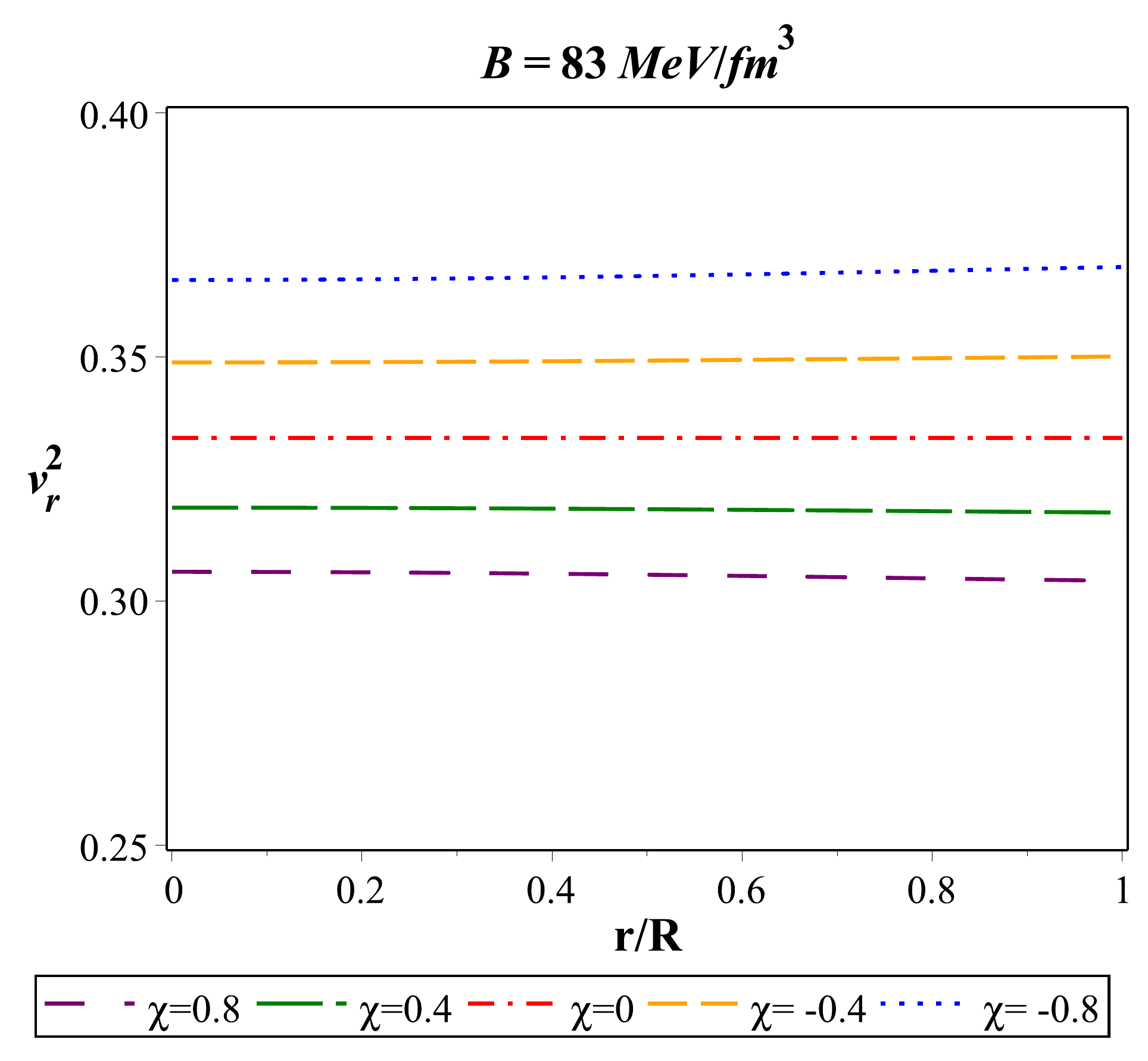}}
    \
    \subfloat{\includegraphics[width=4.5cm]{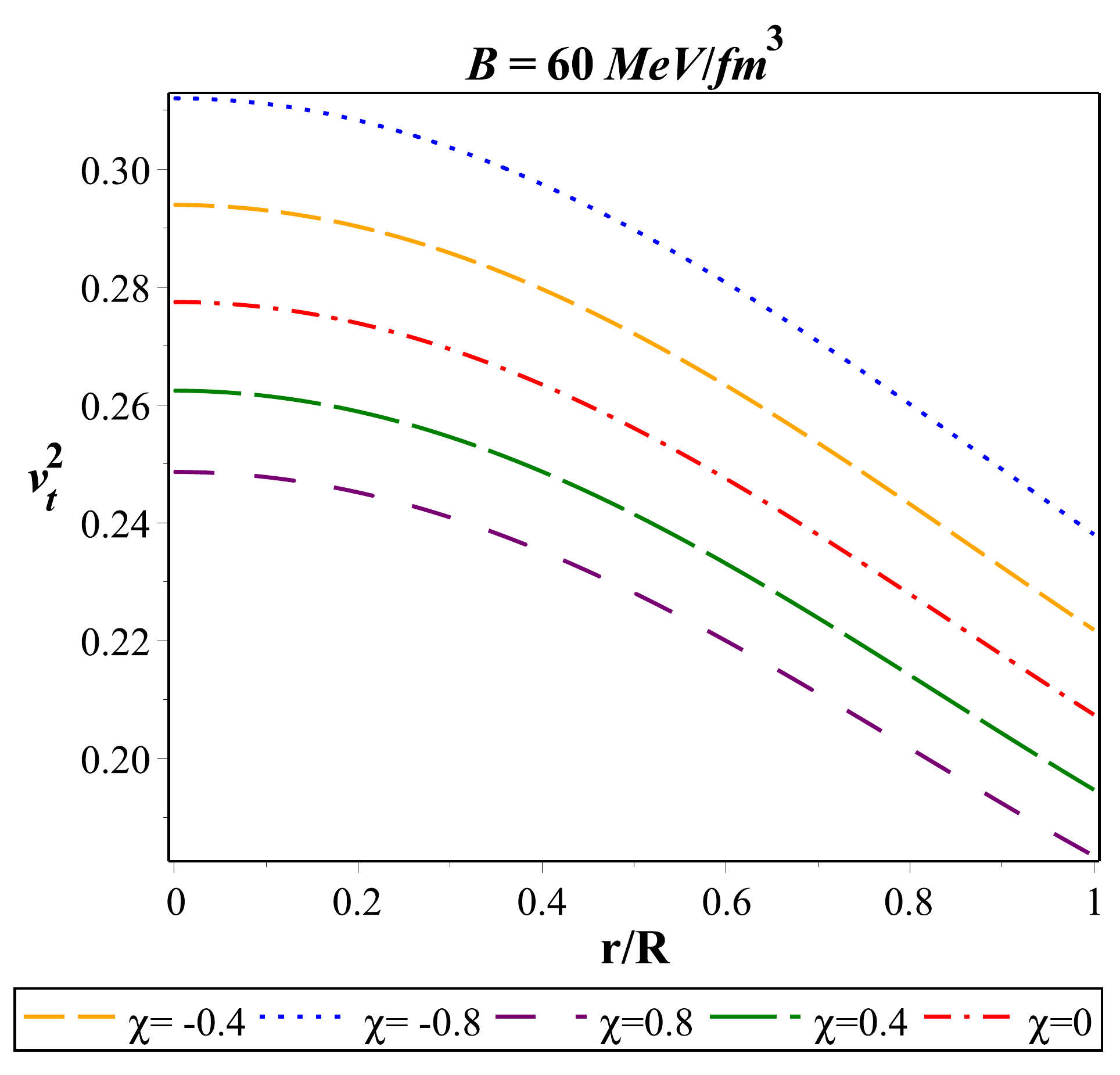}}
    \subfloat{\includegraphics[width=4.5cm]{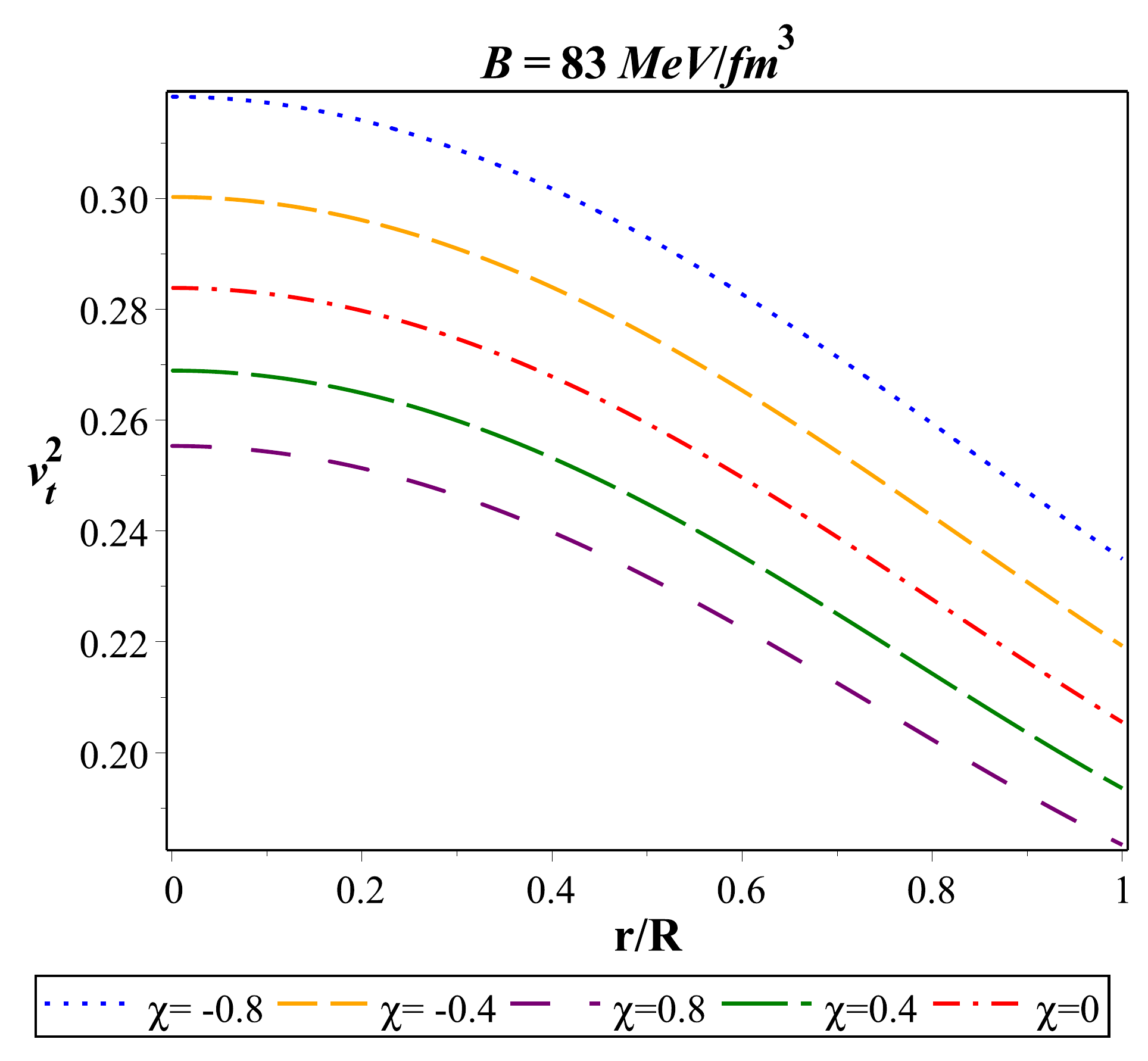}}
    \
    \subfloat{\includegraphics[width=4.5cm]{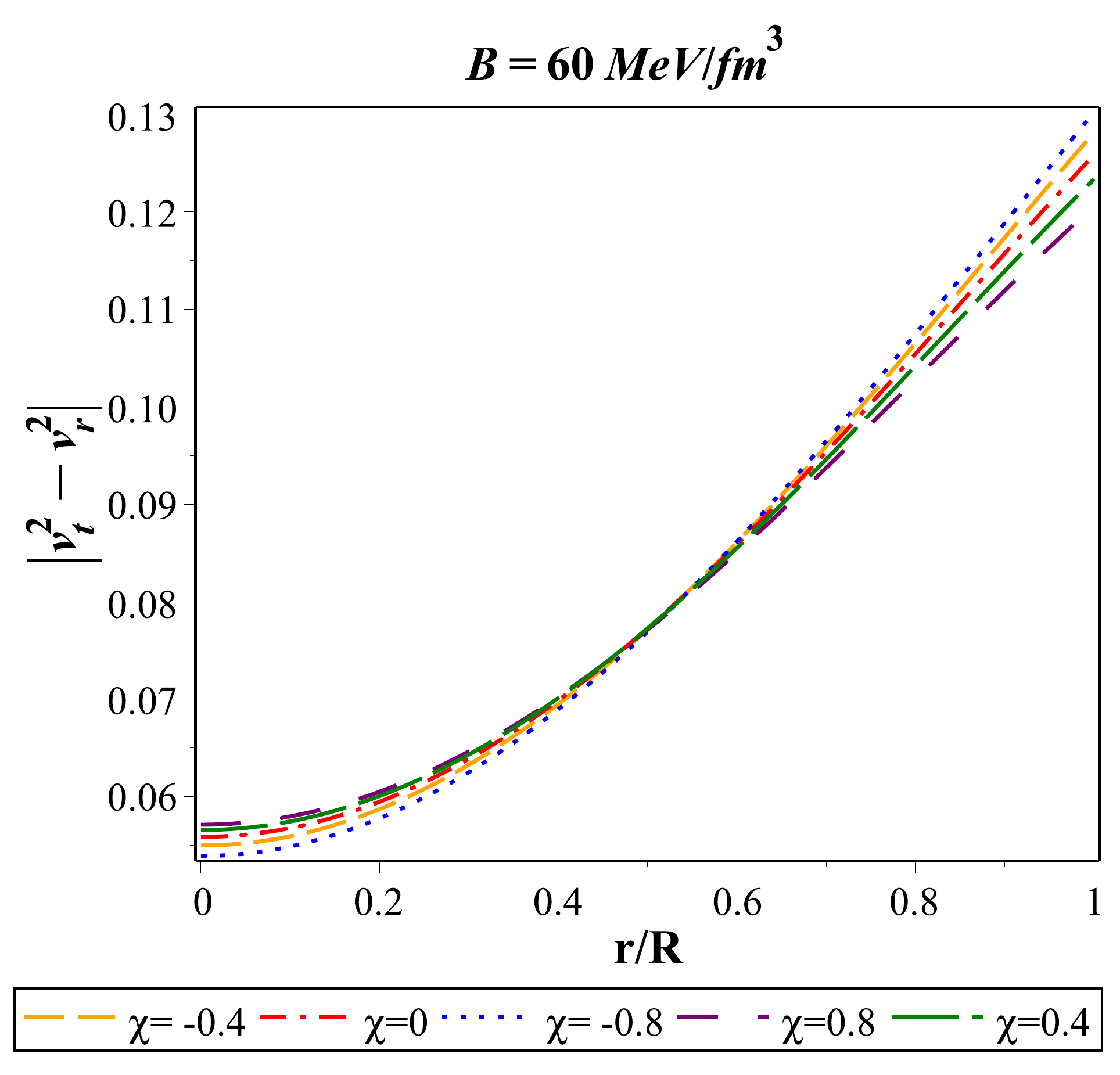}}
    \subfloat{\includegraphics[width=4.5cm]{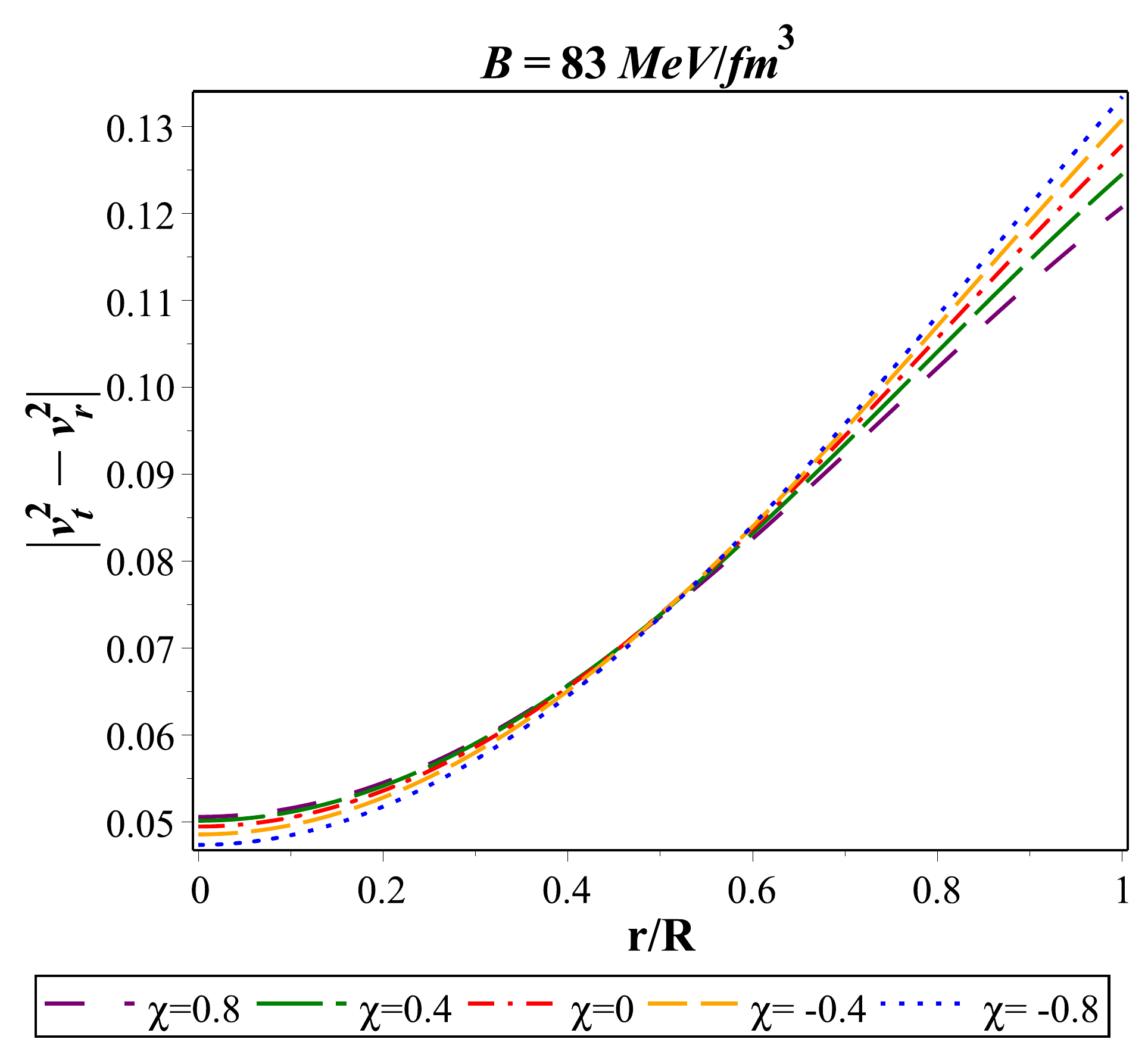}}
\caption{Variation of $v_r^2$ (in the upper panel), $v_t^2$ (in the middle panel) and $\mid v^2_t - v^2_r\mid$ (in the lower panel) with the radial coordinate $r/R$ for $LMC\,X-4$.} \label{Fig7}
\end{figure}

Fig.~\ref{Fig7} shows that our system is consistent with both the causality condition and Herrera cracking concept, which confirms the stability of the system.

\subsubsection{Adiabatic index}\label{subsubsec5.2.3}
For a given density, the stiffness of the equation of states can be characterized by the adiabatic index, and it also describes the stability of both relativistic or non-relativistic compact stars. Following~\citet{Chandrasekhar1964}, dynamical stability of the stellar system against an infinitesimal radial adiabatic perturbation has been tested by many authors~\citep{Hillebrandt1976,Horvat2010,Doneva2012,Silva2015}. In their work~\citet{HH1975} have suggested that for a dynamically stable stellar system the adiabatic index must exceed $\frac{4}{3}$ in all the interior points. For our system, adiabatic index $\Gamma$ reads
\begin{equation}
\Gamma=\frac{p^{\it eff}_r+{\rho}^{\it eff}}{p^{\it eff}_r}\,\frac{dp^{\it eff}_r}{d{\rho}^{\it eff}}=\frac{p^{\it eff}_r+{\rho}^{\it eff}}{p^{\it eff}_r}\,v^2_{r}, \label{5.2.3.1}
\end{equation}

\begin{figure}
\centering
    \subfloat{\includegraphics[width=4cm]{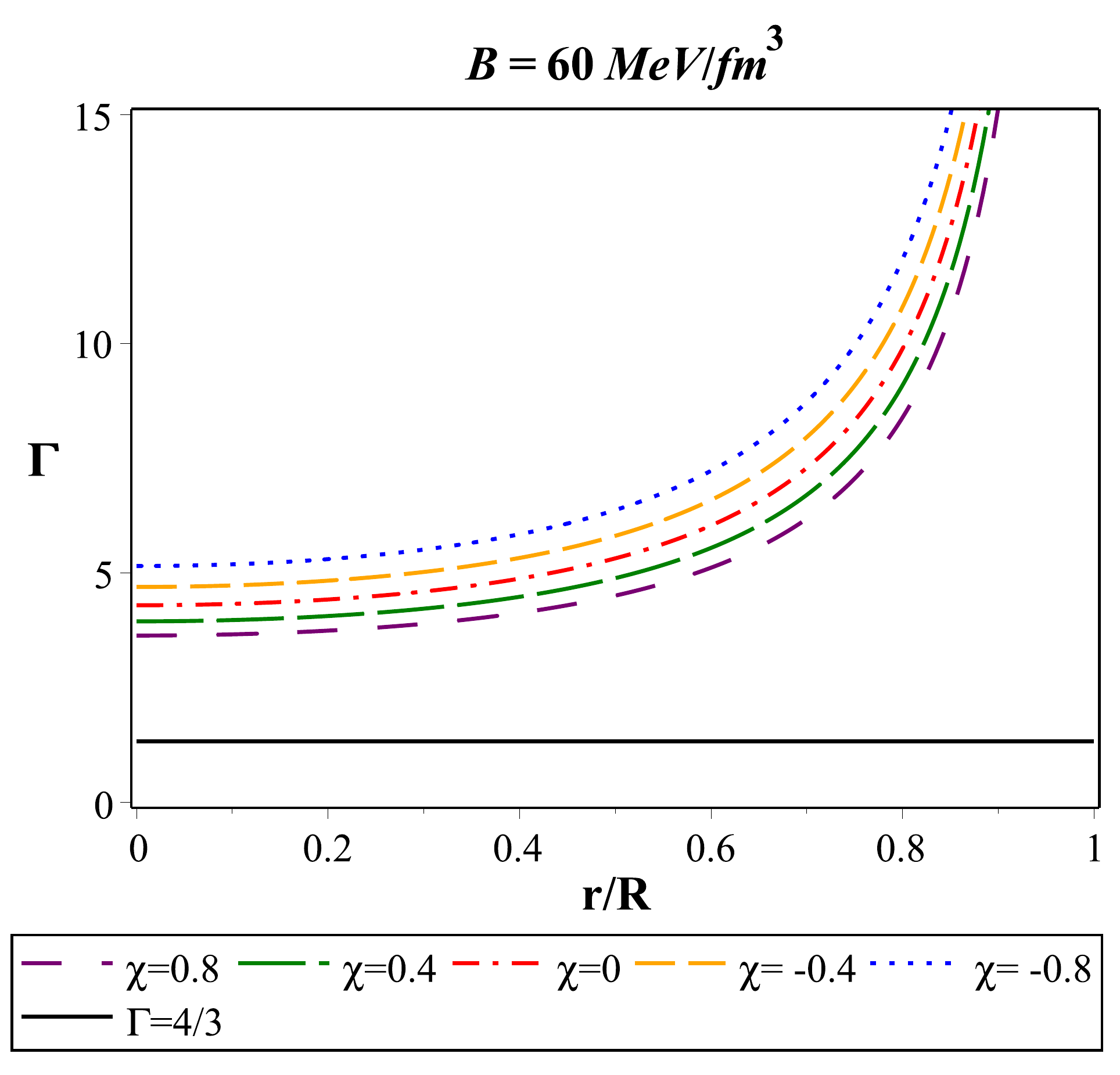}}
    \
    \subfloat{\includegraphics[width=4cm]{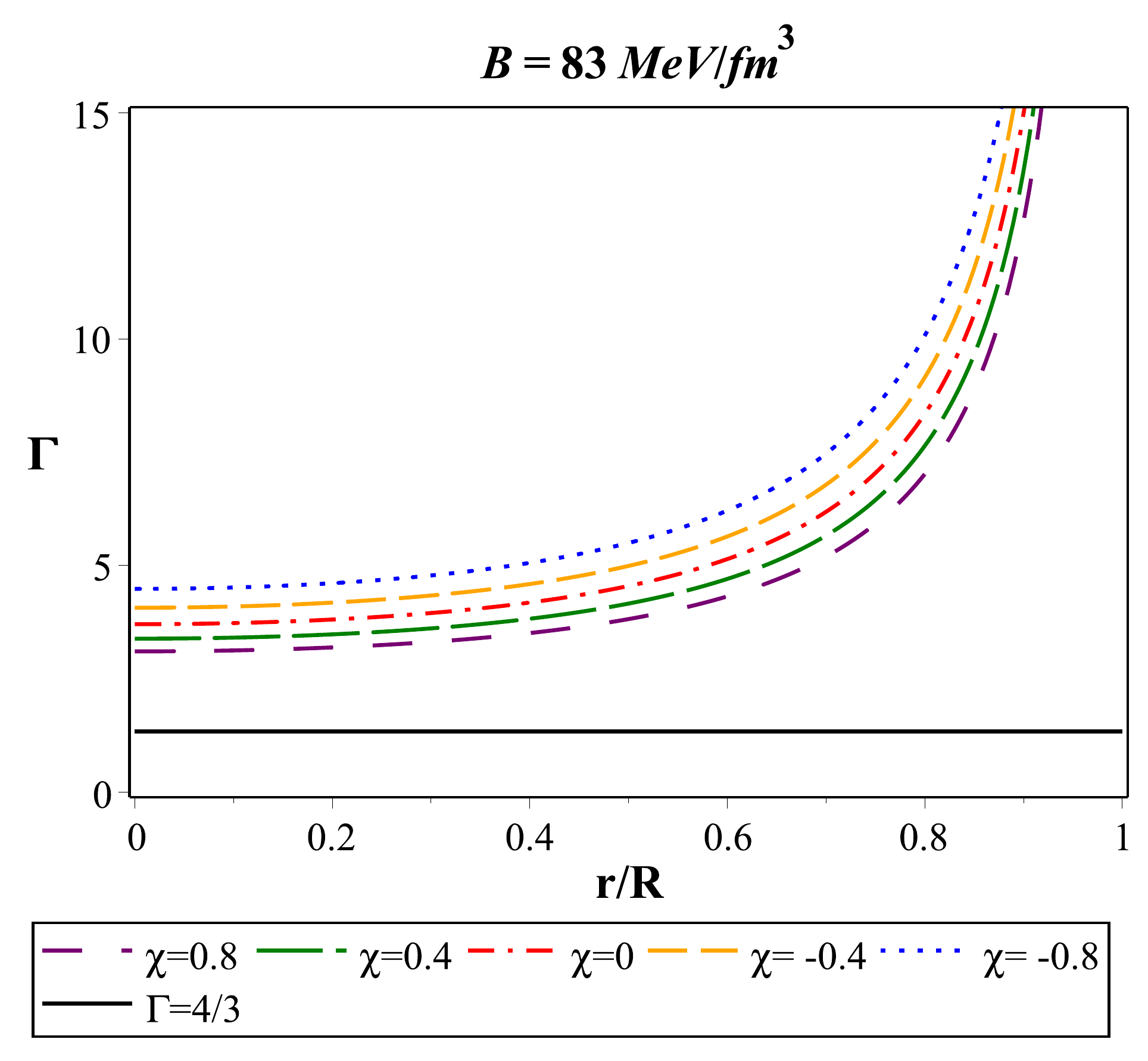}}
    \caption{Variation of adiabatic index $\Gamma$ with the radial coordinate $r/R$ for $LMC\,X-4$.} \label{Fig8}
\end{figure}

The variations of $\Gamma$ against the radial coordinate $r/R$ are presented in Fig.~\ref{Fig8} for $B=60~MeV/{fm}^3$ and $83~MeV/{fm}^3$. It shows that in both cases the adiabatic indices are greater than $\frac{4}{3}$, which confirms the stability of the stellar system.

\subsection{Compactification factor and redshift}\label{subsec5.3}
In our model the mass of the stellar system is given by
\begin{eqnarray}\label{6.1}
&\qquad\hspace{-1cm} m=\frac{1}{2}\left[-\frac{1}{1-\frac{2\,M{r}^{2} \left\lbrace -{\frac {M{r}^{2}}{{R}^{2} \left( 2\,Mn-Rn+M \right) }}+1 \right\rbrace ^{n-2}}{\left\lbrace {\frac {n \left( -R+2\,M \right) }{2\,Mn-Rn+M}} \right\rbrace ^{n-2} \left( -R+2\,M \right) {R}^{2}}}+1\right] r. 
\end{eqnarray}

From Eq.~(\ref{6.1}) we find that the mass function is regular within the stellar system and is zero at the centre, when $r=0$.

The compactification factor is defined by
\begin{eqnarray}\label{6.2}
&\qquad\hspace{-6.82cm} u(r)=\frac{m(r)}{r}\nonumber \\
&\qquad\hspace{-0.5cm} =\frac{1}{2}\left[-\frac{1}{1-\frac{2\,M{r}^{2} \left\lbrace -{\frac {M{r}^{2}}{{R}^{2} \left( 2\,Mn-Rn+M \right) }}+1 \right\rbrace ^{n-2}}{\left\lbrace {\frac {n \left( -R+2\,M \right) }{2\,Mn-Rn+M}} \right\rbrace ^{n-2} \left( -R+2\,M \right) {R}^{2}}}+1\right].
\end{eqnarray}

\begin{figure}
\centering
    \subfloat{\includegraphics[width=4.5cm]{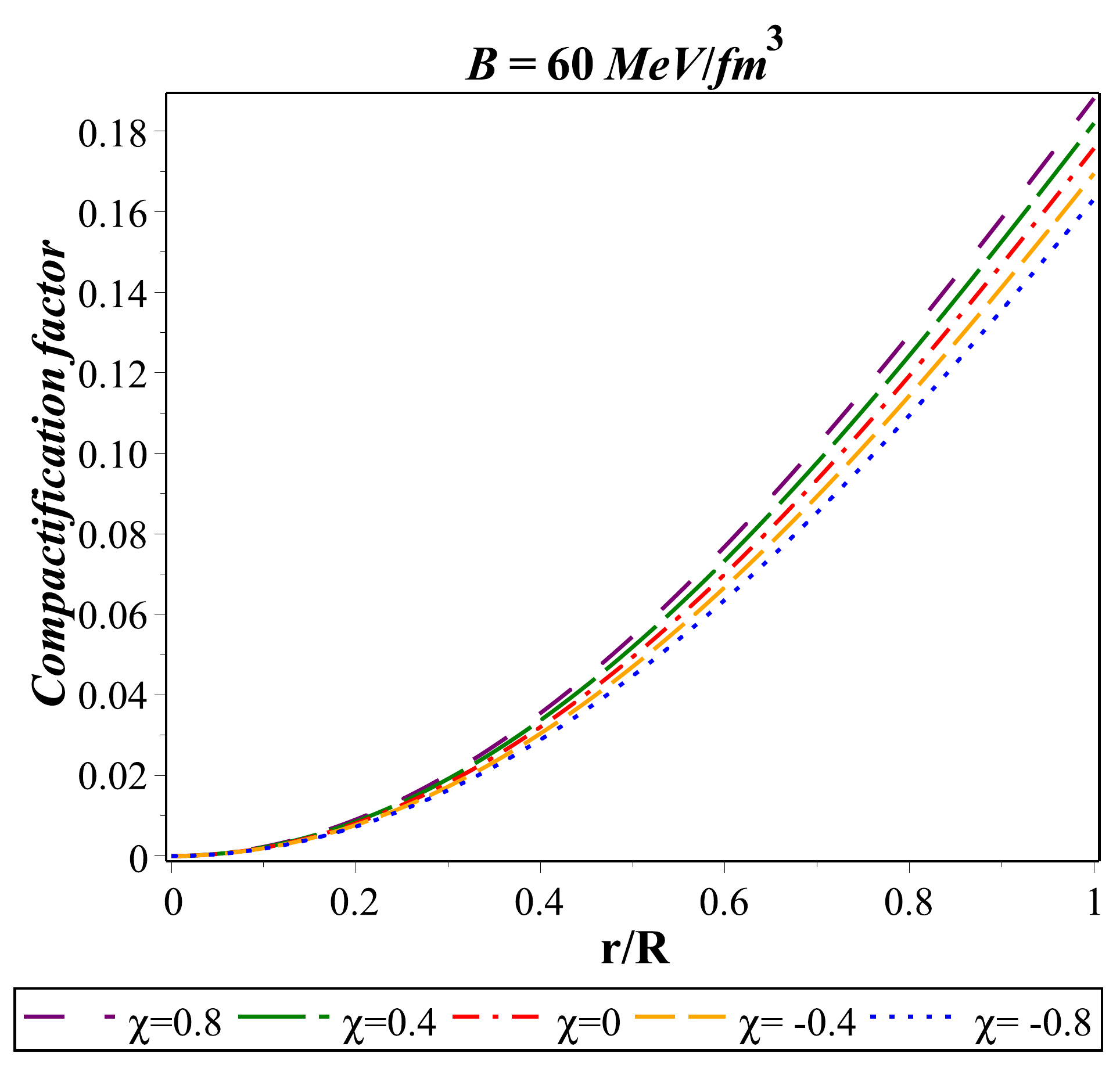}}
    \subfloat{\includegraphics[width=4.5cm]{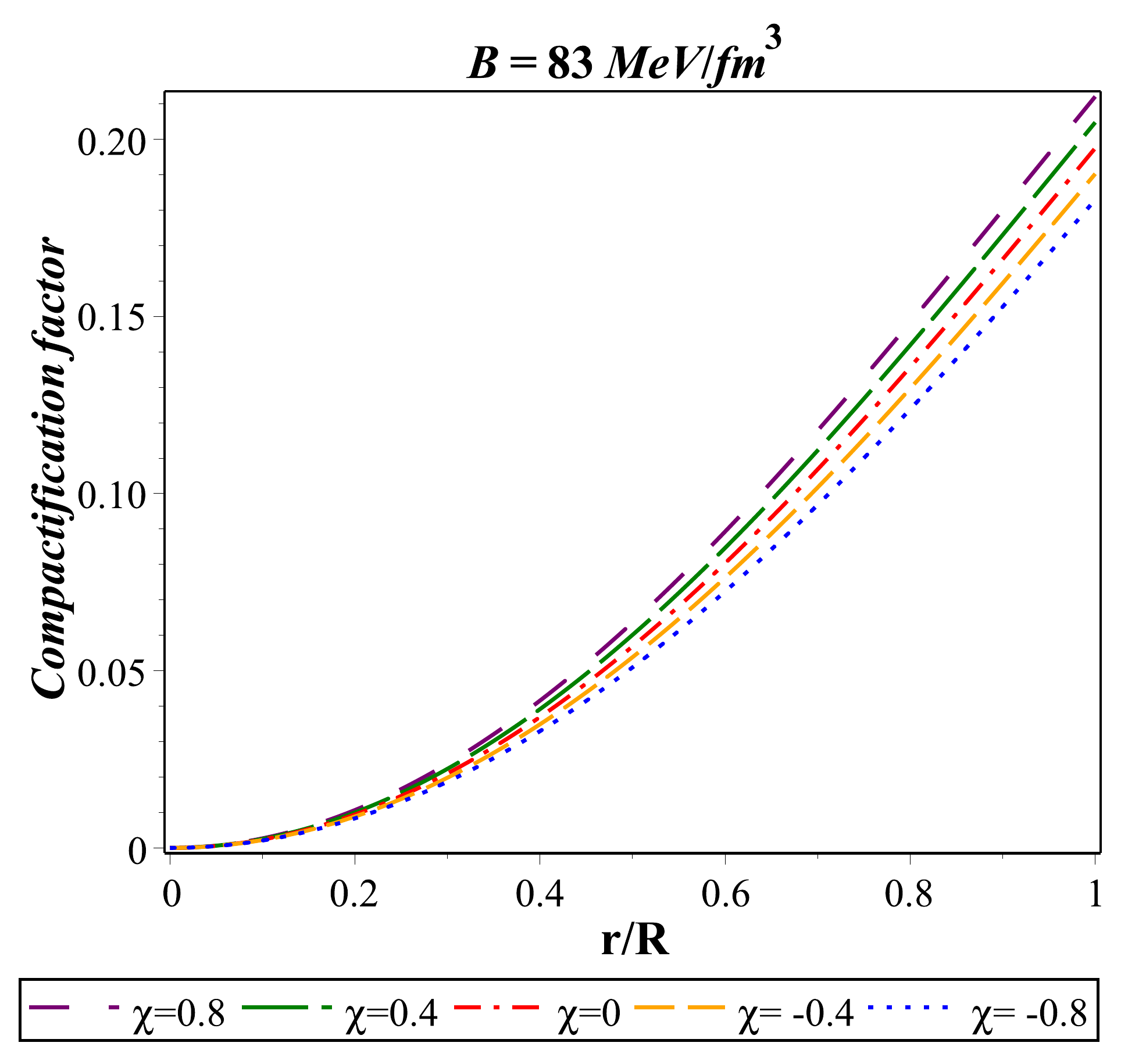}}
        \caption{Variation of compactification factor with the radial coordinate $r/R$ for $LMC\,X-4$.} \label{Fig9}
\end{figure}

The variation of the compactification factor is shown in Fig.~\ref{Fig9}. Note that for a spherically symmetric stellar system the maximally allowed mass-to-radius ratio, as predicted by~\citet{Buchdahl1959}, is given by $\frac{M}{R}<\frac{4}{9}$. Fig.~\ref{Fig9} shows that for all the values of $\chi$ the Buchdahl condition is satisfied in our system. 

The surface redshift for our system is defined by
\begin{equation}\label{6.3}
Z=\frac{1}{\sqrt{\frac{\left( R-2\,M \right)  \left\lbrace 1-{\frac {M{r}^{2}}{{R}^{2} \left( 2\,M
n-Rn+M \right) }} \right\rbrace ^{n}}{\left\lbrace {\frac {n \left( -R+2\,M \right) }{2\,Mn-Rn+M}} \right\rbrace ^{n}R}}}-1.
\end{equation}

\begin{figure}
\centering
    \subfloat{\includegraphics[width=4.5cm]{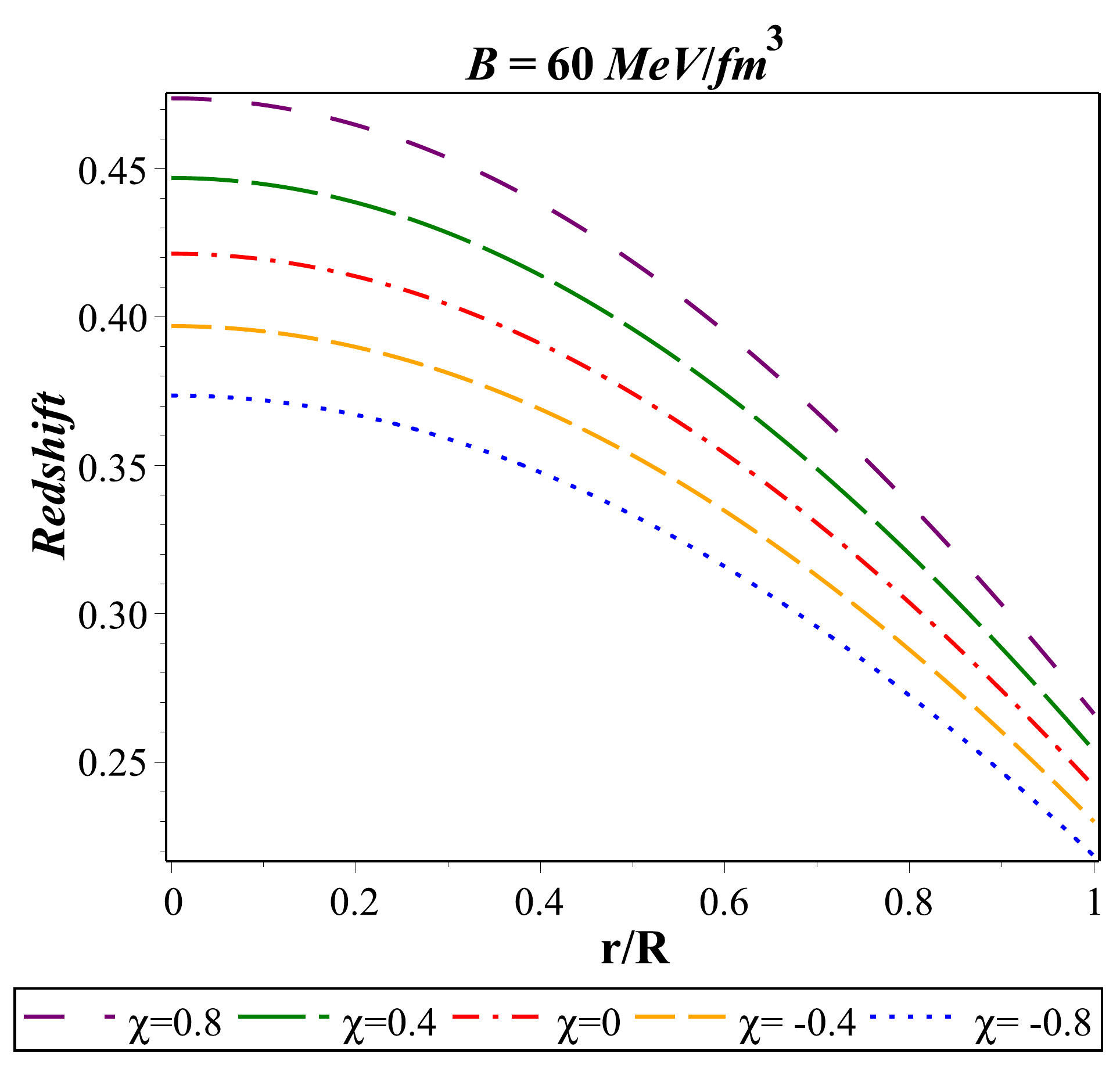}}
    \subfloat{\includegraphics[width=4.5cm]{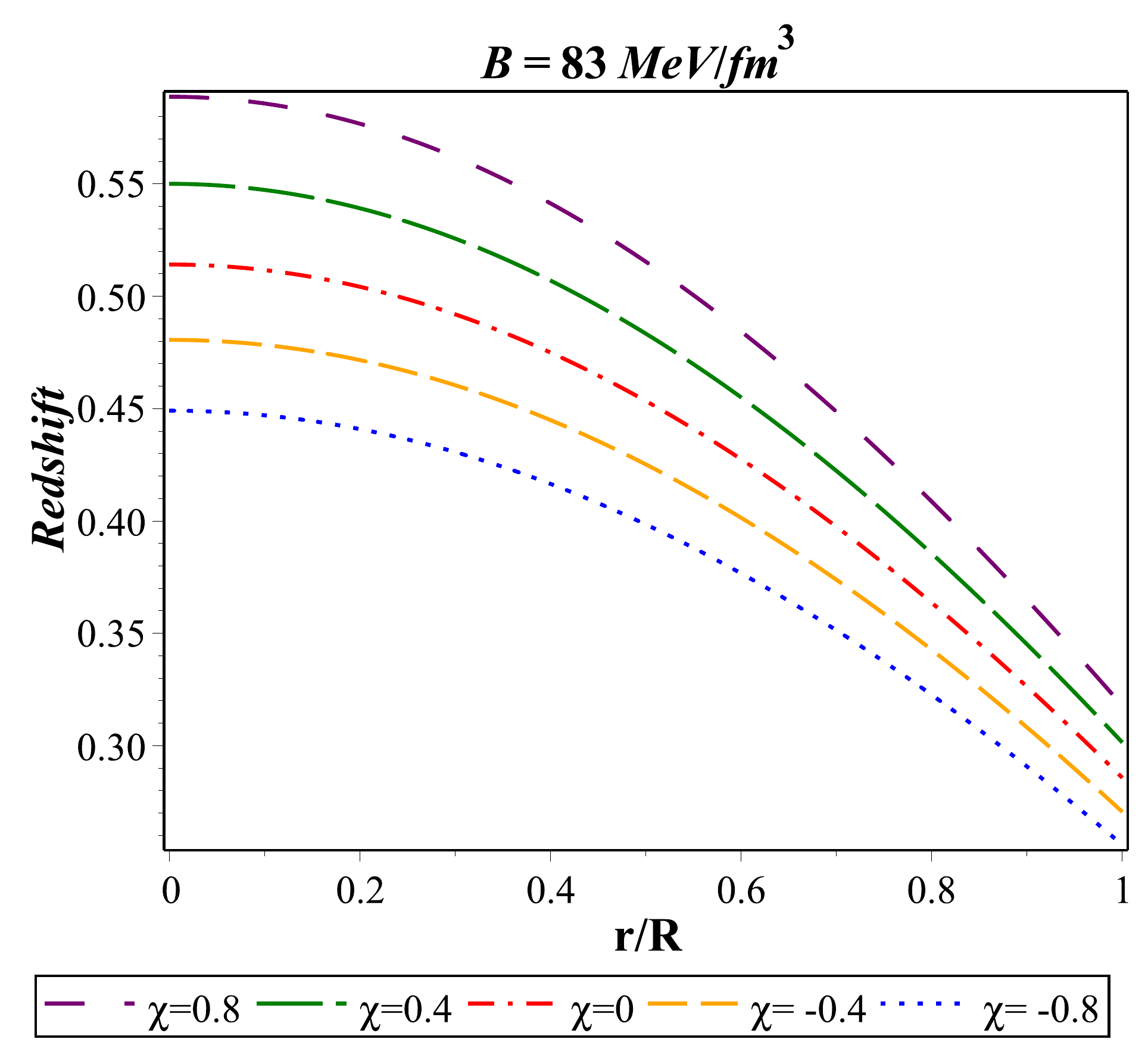}}
\caption{Variation of redshift with the radial coordinate $r/R$ for $LMC\,X-4$.} \label{Fig10}
\end{figure}

The variation of the redshift with the radial coordinate is shown in Fig.~\ref{Fig10}.

\section{Discussion and conclusion}\label{sec6}
Our investigation is devoted to a new class of generalized solutions for the spherically symmetric anisotropic compact stellar objects, especially strange stars. In our study, we corroborate the `embedding class' method as a powerful instrument to obtain general solutions of the modified Einstein field equations in the framework of the $f\left(R,\mathcal{T}\right)$ gravity theory. To be precise, our study deals with the `embedding class 1' method where a four-dimensional interior spacetime is embedded into the five-dimensional flat Euclidean space. To incorporate the interaction between the geometrical terms and matter, following~\citet{harko2011} we considered the simplified (conservative) linear ansatz for the $f\left(R,\mathcal{T}\right)$ function given by $f\left(R,\mathcal{T}\right)=R+h\left(T\right)$. This modification results in the modification of the standard Einstein field equations, as presented in Eq.~\ref{1.5}, and indicates the presence of a new type of matter inside the stellar system along with SQM. This new type of matter originates due to the specific interaction between the geometric term and matter~\citep{SC2013}. Equation (\ref{1.6}) features the non-conservation of the energy-momentum tensor, but this can be resolved by considering the effective energy-momentum tensor of the matter distribution as $T^{\it eff}_{\mu\nu}$~in Eq.~(\ref{1.5a}) that results in $\nabla^{\mu}T^{\it eff}_{\mu\nu}=0$. To solve the gravitational field equations~(\ref{2.2})-(\ref{2.4}), we have applied the embedding class one techniques (\ref{3.2}) and considered the SQM to be governed by the phenomenological and simplified MIT bag EOS~(\ref{2.5}). Following~\citet{Lake2003}, we assumed the time-time component of the metric potential to be  ${{\rm e}^{\nu}}=G \left( A{r}^{2}+1 \right) ^{n}$ and obtained ${\rm e}^{\lambda}$ in Eq.~(\ref{3.4}).

For the numerical analysis and study of the physical justifications of the achieved solutions, we consider $LMC~X-4$ as the representative of the strange star candidates, having the observed mass as $1.29$~$M_{\odot}$~\citep{Rawls2011}. Further, to derive the unknown values of the constants, viz., $n$, $A$, $C$, $F$, and $G$, etc., and radius $R$ of the stellar system, we assumed random values of $\chi$ as $-0.8$, $-0.4$, $0$, $0.4$ and $0.8$. We also assumed the values of the bag constant $B$ as $60~MeV/{fm}^3$ and $83~MeV/{fm}^3$ as the examples, which lie well within the accepted values of $B$~\citep{Burgio2002,Kalam2013,Rahaman2014}. To avoid confusion, it should be mentioned that there is no specific reason behind the choice of our values for $\chi$ and $B$: their values were chosen randomly, in order to derive unknown parameters of the stellar system. 

The behaviour of the metric potentials, viz., ${\rm e}^{\nu}$ and ${\rm e}^{\lambda}$ are shown in Fig.~\ref{Fig1}, for the chosen values of the bag constants. The variation of the effective matter distribution ${\rho}^{\it eff}$ against the radial coordinate $r/R$ is shown in Fig.~\ref{Fig2}. The behaviour of the effective radial $(p^{\it eff}_{r})$ and tangential~$(p^{\it eff}_{t})$ pressures are featured in the upper and lower panel of Fig.~\ref{Fig3}. Importantly, Figs.~\ref{Fig1} - \ref{Fig3} exhibit that our system is completely free from any type of singularities, viz., geometrical and physical singularities. In Fig.~\ref{Fig4} we have shown the behaviour of the anisotropic stress of the system, which is zero at the centre and reaches its maximum at the surface. It is interesting to note that in the framework of $f\left(R,\mathcal{T}\right)$ gravity theory the anisotropy follows the same behaviour as predicted by~\citet{Deb2017} in the case of GR. In Fig.~\ref{FigMR} we present the mass-to-radius relation for the strange star candidates, that appears to be similar to the typical $M-R$ curve for the strange star candidates in GR. We find for $B=83~MeV/{fm}^3$, as the values of $\chi$ decrease from $\chi=0.8$ to $\chi=-0.8$, the values of maximum mass $M_{max}$ rise $25.362~\%$, whereas the corresponding values of total radius $\tilde{R_{M}}$ rise $26.819~\%$. In each case, we also calculated the values of the constant $n$, which increase $0.335~\%$ when $\chi$ decreases from $\chi=0.8$ to $\chi=-0.8$. In summary, when the values of $\chi$ decrease, the compact stellar objects become more massive and larger in size, which turns them into less dense stellar objects gradually.

Figure~\ref{Fig5} features that our system is consistent with all the energy conditions, and it confirms the physical validity of the achieved solutions. To verify whether all the forces are in equilibrium for our system, we studied the modified TOV equation for $f\left(R,\mathcal{T}\right)$ gravity. Fig.~\ref{Fig6} shows that all the forces are in equilibrium, which confirms the stability of the system. Interestingly, a new kind of force $F_m$ is arising in our system due to the coupling between matter and geometry. This force is repulsive in nature and acts along the outward direction when $\chi<0$, whereas $F_m$ acts along the inward direction and shows the attractive nature for $\chi>0$. The upper and middle panel of Fig.~\ref{Fig7} shows that the square of the sound speeds, viz., $v^2_r$ and $v^2_t$ are well within the interval between $0$ and $1$, and the present system is consistent with the causality condition. Again, from the lower panel of Fig.~\ref{Fig7} we get that our system is also consistent with the Herrera cracking condition, as we have $0<\mid v^2_{t} - v^2_{r}\mid <1$, which assures the stability of the present anisotropic stellar system. Figure~\ref{Fig8} features that in all the cases our system is stable against an infinitesimal radial adiabatic perturbation, as $\Gamma$ is greater than $4/3$ in all interior points of the stellar configuration. We have also presented the behaviour of the compactification factor and the redshift function in Figs.~\ref{Fig9} and \ref{Fig10}, respectively.

For $\chi=-0.4$ and $B=83~MeV/{fm}^3$ we have predicted values of various physical parameters, viz., the radius ($R$), central density ($\rho^{\it eff}_c$), surface density ($\rho^{\it eff}_0$), central pressure ($p_c$), mass-radius ratio ($2M/R$) and surface redshift ($Z_s$) in Table~\ref{Table 1}, and have also presented the calculated values of the different constants, viz., $\chi$, $A$, $C$ and $G$ in Table~\ref{Table 2}, due to different strange star candidates.  Table~\ref{Table 1} features that the strange star candidates have high redshift values~$(0.182-0.449)$, and their surface densities~$(2.338\rho_0 - 2.343\rho_0)$ are higher than the normal nuclear density ${\rho}_0$~\citep{Ruderman1972,Glendenning1997,Herzog2011}, which confirms that the chosen stellar configurations are suitable candidates for the hypothetical ultra-dense strange stars. Further, with the motivation for the comparative discussion, in Table~\ref{Table 3} we have presented the predicted values of the physical parameters for the strange star candidate $LMC~X-4$ due to chosen parametric values of $\chi$, viz., $\chi=-0.8, -0.4, 0, 0.4$ and $0.8$, and the bag constants $B=83 MeV/{fm}^3$. Interestingly, Table~\ref{Table 4} features that as values of $B$ increase the values of $R$ and $n$ decrease accordingly, however, the values of the physical parameters $\rho^{\it eff}_c$, $\rho^{\it eff}_0$, $p_c$, $2M/R$ and $Z_s$ increase gradually. Hence, Tables~\ref{Table 3} and \ref{Table 4} clearly show that as the values of $\chi$ and $B$ decrease, the stellar systems become larger in size turning them into less dense compact objects.

It is interesting to mention that our study on the effect of the $f\left(R,\mathcal{T}\right)$ gravity theory on the strange star candidates reveals the same result as predicted by~\citet{Astashenoka2015} in their study, which deals with a non-perturbative models of strange stars in $f\left(R\right)=R+\alpha {R}^2$ theory of gravity, where $\alpha$ is a constant parameter. Their study~\citep{Astashenoka2015} reveals that as the values of $\alpha$ increase, the stellar system becomes more massive gradually, which in our study is an interesting case due to the decreasing values of the constant parameter $\chi$. The authors~\citep{Astashenoka2015} predicted that the presence of the extra geometrical term $\alpha R^2$ is the reason behind the origin of the extra gravitational mass, whereas in our study it is the case but due to the presence of the extra matter term $2\chi\mathcal{T}$. In our study, we find in the framework of the particular $f\left(R,\mathcal{T}\right)$ gravity theory, as the values of $\chi$ increase, the values of surface redshift ($Z_s$) increase gradually, whereas in the case of $f\left(R\right)$ gravity~\citet{Astashenoka2015} found that $Z_s$ decrease consequently with the increase of the values of $\alpha$. Hence, the reader may easily differentiate the effects of the $f\left(R,\mathcal{T}\right)$ and $f\left(R\right)$ gravity by analyzing the surface redshift of compact stellar objects.

Einstein's theory has been tested successfully mainly in the regime of weak gravity through solar system tests and laboratory experiments. But the validity of this highly accepted theory still faces stringent constraint in the regime of strong gravity, viz., the region near to a black hole, ultra dense compact stars and expanding universe. The recent discovery of peculiar highly over-luminous SNeIa, e.g. SN 2003fg, SN 2006gz, SN 2007if and SN 2009dc~\citep{Howell2006,Scalzo2010} indicates a huge Ni-mass and confirms the highly super-Chandrasekhar white dwarfs, having mass $2.1-2.8~M_{\odot}$, as a suitable progenitors~\citep{Howell2006,Scalzo2010,Hicken2007,Yamanaka2009,Silverman2011,Taubenberger2011}. Recently, \citet{Linares2018} have discovered a highly massive pulsar of mass $2.27^{+0.17}_{-0.15}~M_{\odot}$ in their observation of compact binaries $PSR~ J2215+5135$. Clearly, these observations are not only questioning the standard Chandrasekhar limit for the compact stellar objects but also invoking the necessity of modification of GR in the strong gravity regime. Interestingly, our study reveals that due to the effect of $f\left(R,\mathcal{T}\right)$ gravity theory, the maximal mass limits rise higher than their standard values in GR for the chosen parametric values of $\chi$. Hence, the stellar systems in the framework of $f\left(R,\mathcal{T}\right)$ gravity theory may also explain the observed massive stellar systems, viz., massive pulsars, super-Chandrasekhar stars and magnetars, etc., which GR hardly can explain suitably so far. In support of the achieved result in the present investigation, it is worth mentioning that the important study by~\citet{Laurentis2018} also reveals that the application of the Noether Symmetry Approach can explain the extreme massive stars which supposed to be unstable in the framework of GR. However, in the limit $\chi=0$, one may retrieve the solutions of the standard Einstein gravity.

As a final comment, by applying the `embedding class 1' techniques, we successfully developed a singularity free and stable anisotropic generalized model, suitable for studying the ultra-dense strange stars in the framework of $f\left(R,\mathcal{T}\right)$ gravity theory.

\section*{Acknowledgments}

SR is thankful to the Inter-University Centre for Astronomy and Astrophysics (IUCAA), Pune, India and the Institute of Mathematical Sciences, Chennai, India for providing Visiting Associateship under which a part of this work was carried out. Another part of this work was completed while DD was visiting the IUCAA, and DD gratefully acknowledges the warm hospitality received there. The work by SVK on $f(R, T)$ gravity was supported in part by the Competitiveness Enhancement Program of Tomsk Polytechnic University in Russia, by a Grant-in-Aid of the Japanese Society for Promotion of Science (JSPS) under No. 26400252, and the World Premier International Research Center Initiative (WPI Initiative), MEXT, Japan. SVK is grateful to the Institute for Theoretical Physics at Vienna University of Technology in Austria for kind hospitality extended to him during part of this investigation. The work by PM on strange stars is supported by the grand FAPESP (Funda{\c c}{\~ a}o de Amparo {\'a} Pesquisa do Estado de S{\~ a}o Paulo), grant 2015/08476-0. The work by MK was supported by the Ministry of Education and Science of the Russian Federation, MEPhI Academic Excellence Project (contract 02.a03.21.0005, 27.08.2013). One of the authors (DD) is thankful to Tiberiu Harko for his pertinent suggestions which helped to upgrade the manuscript substantially. We are also thankful to the anonymous referee for drawing our attention to some important and relevant works which helped us to upgrade this manuscript.

\end{document}